\documentclass{article}

\usepackage{arxiv}

\usepackage[utf8]{inputenc} 
\usepackage[T1]{fontenc}    
\usepackage{hyperref}       
\usepackage{url}            
\usepackage{booktabs}       
\usepackage{amsmath, amssymb, amsfonts}       
\usepackage{nicefrac}       
\usepackage{microtype}      
\usepackage{lipsum}		
\usepackage{graphicx}
\usepackage{natbib}
\usepackage{doi}
\usepackage{graphicx}
\usepackage{subfigure}
\usepackage{geometry}
\usepackage{textcomp}
\usepackage{multirow} 
\usepackage{multicol} 
\usepackage{nomencl}
\usepackage{framed}

\title{Aerodynamic characterization of two tandem wind turbines under yaw misalignment control using actuator line model}

\author{ Jie Su \\
	School of Naval Architecture, Ocean and Civil Engineering\\
	Shanghai Jiao Tong University\\
	Shanghai 200240, China \\
	\And
  	Yu Tu \\
	School of Naval Architecture, Ocean and Civil Engineering, Shanghai Jiao Tong University \\
	Shanghai 200240, China \\
	\And
 	Limin Kuang \\
	School of Naval Architecture, Ocean and Civil Engineering, Shanghai Jiao Tong University \\
	Shanghai 200240, China \\
	\And
 	Rui Zhang \\
	School of Naval Architecture, Ocean and Civil Engineering, Shanghai Jiao Tong University \\
	Shanghai 200240, China \\
	\And
 	Yixiao Shao  \\
	School of Naval Architecture, Ocean and Civil Engineering, Shanghai Jiao Tong University \\
	Shanghai 200240, China \\
	\And
 	Dai Zhou \\
	School of Naval Architecture, Ocean and Civil Engineering, Shanghai Jiao Tong University \\
	Shanghai 200240, China \\
	  \texttt{zhoudai@sjtu.edu.cn} \\
	\And
 	Zhaolong Han \\
	School of Naval Architecture, Ocean and Civil Engineering, Shanghai Jiao Tong University \\
	Shanghai 200240, China \\
	\And
        \href{https://orcid.org/0000-0001-6097-7217}
	{\includegraphics[scale=0.06]{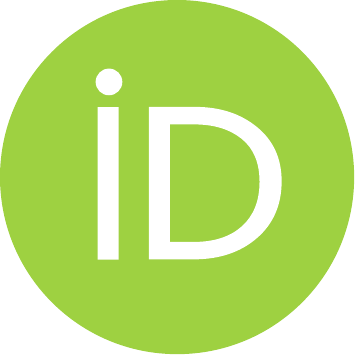}\hspace{1mm}Kai Zhang} \\
	School of Naval Architecture, Ocean and Civil Engineering\\
	Shanghai Jiao Tong University\\
	Shanghai 200240, China \\
}

\date{February 17,2023}

\renewcommand{\shorttitle}{}

\begin{document}
\maketitle

\begin{abstract}
The Darrieus vertical axis wind turbines (VAWTs) are one of the mainstream devices for wind energy utilization in the urban areas. The market's pursuit of high wind energy conversion efficiency promotes the research on improving the wind turbine power coefficient while reducing noise. This paper makes further investigation on the aerodynamics and aeroacoustics of a small VAWT with V-shaped blades and trailing-edge serrations. The feasibility of utilizing the Reynolds-Averaged Navier-Stokes SST $k-\omega$ turbulence model and the FW-H method is verified against experiments. The studied V-shaped blades can effectively improve the power performance of VAWT over a wide range of tip speed ratios under normal wind speed conditions, and the trailing-edge serrations will also slightly increase the power output of V-bladed VAWT at the optimal tip speed ratio. The power coefficient of the V-bladed wind turbine with trailing-edge serrations is about 28.3$\%$ higher than that of the original turbine. In addition, a dumbbell-shaped noise directivity distribution was first discovered in the VAWT compared with the traditional elliptical distribution. The V-bladed VAWT generated less low-frequency noise and the trailing-edge serrations realized the expected noise reduction effect. Practically, this study proposes a feasible solution for the design of high-efficiency and low-noise wind turbines.
\end{abstract}


\section{Introduction}
\label{sec:intro}
Due to the severe climate crisis and environmental pollution, reducing carbon emissions has become the common goal of the government around the world. Wind energy is considered one of the most important sustainable and green energy sources and plays an important role in the supplementation of electricity in urban and residential areas. Due to the density of buildings, complex topographical conditions and unique wind conditions in urban areas, the small-scale vertical axis wind turbines (VAWTs) are becoming increasingly popular in the urban environment \cite{li2022experimental}. 

Compared with the horizontal axis wind turbines (HAWTs), the Darrieus VAWTs are one of the mainstream devices for wind energy utilization in the urban areas \cite{griffith2016study}. However, the relatively lower aerodynamic efficiency and incomplete corollary equipment impede the rapid development of VAWTs. Besides, the public still has different attitudes towards wind turbine noise. Therefore, it is urgent to further study and evaluate the aerodynamic and aeroacoustic performance of VAWT to overcome its defects.

\subsection{Improvement of VAWT’s structure}

In order to increase the power output and improve the cost performance of VAWT, in recent years, researchers have made great efforts to improve the VAWT’s structure, including the structural configuration and the blade profile. 

In the early stage of development of Darrieus VAWT, the subject of study is generally curve-bladed turbines, namely the $\Phi$-type VAWT. After that, to reduce manufacturing cost of blade and simplify the support structure, the straight-bladed (H-type rotor) VAWT was largely developed in the UK \cite{borg2014offshore}. Due to its simple structure, clear structural force transmission path, and easy fabrication and installation, this H-type Darrieus VAWT has been used as a baseline model for numerous experimental and numerical studies. Besides, to improve the power performance and reduce the power output fluctuations, the helically twisted configuration inspired by Gorlov hydrokinetic turbines was introduced in VAWTs. Scheurich and Brown \cite{scheurich2013modelling} investigated the behavior of helical VAWT. The helical turbine was shown to experience lower power loss when blade curvature and helical twist were optimized, and it would generate more power output than the straight-bladed VAWT at constant rotational speed under unsteady wind conditions. Marten et al. \cite{marten2019benchmark} performed an aero-elastic simulation for a helical VAWT, and the aerodynamic performance was evaluated by the lifting line free vortex wake model, which was validated against experimental data. By eliminating the horizontal struts and supporting arms of turbine rotor, Xu et al. \cite{xu2018experimental} conducted an experimental and numerical investigation on a V-rotor wind turbine which mimics the lower half of the $\Phi$-type Darrieus VAWT. The results showed that a higher power coefficient could be obtained by optimizing the pitch angle and opening angle. On the other hand, based on the H-type VAWT, Arpino et al. \cite{arpino2018numerical} investigated a novel VAWT with three couples of blades in Computational Fluid Dynamics (CFD) simulations. The blade group consisting of one main blade and one auxiliary blade was reported to make higher power output at low tip speed ratios compared with that of traditional H-type wind turbine. Liu et al. \cite{liu2019enhancing} integrated the Savonius wind turbine and H-type Darrieus wind turbine to improve the self-starting capability of VAWT. In addition, Kuang et al. \cite{kuang2022wind} and Chen et al. \cite{chen2021efficiency} analyzed the effects of diffuser and deflector on the VAWT. The results showed that the power efficiency was improved by 20$\%$ and 51$\%$ by utilizing a suitable deflector and the optimized diffuser, respectively.

As for the modification in blade profile employed in VAWTs, Ismail and Vijayaraghavan \cite{ismail2015effects} utilized the Shear Stress Transport (SST) $k-\omega$ model to investigate the effects of inward semi-circular dimple and Gurney flap on the power performance of a VAWT. It was found that the average tangential force of the rotor was increased by about 35$\%$ in steady state case and 40$\%$ in oscillating case. Similarly, Zhu et al. \cite{zhu2021effect} evaluated the aerodynamic performance of VAWT with different Gurney flaps, and the maximum improvement could reach up to about 21$\%$. Besides, Tiranda and Rezaeiha \cite{tirandaz2021effect} performed an analysis of airfoil maximum thickness and its position, as well as the leading-edge radius to identify the optimal blade profile for designing VAWTs with morphing airfoils. When tip speed ratio reduced from 3 to 2.5, the optimal maximum thickness increased from 18$\%c$ to 24$\%c$, and its position shifted from 27.5$\%c$ to 35$\%c$. In addition, Zamani et al. \cite{zamani2016starting} proposed a J-shaped profile based on Du 06-W-200 airfoil to improve the self-starting of a VAWT, and the performance of turbine was optimized. According to this idea, Mohamed \cite{mohamed2019criticism} numerically investigated the aerodynamic and aeroacoustic performance of J-shaped Darrieus VAWT. However, the results showed that the studied J-shaped blade has no benefit for wind turbine power performance and noise reduction. 

\subsection{Influence of noise emitted from wind turbines}
Due to the longer development time and a more complete industrial chain, the large-scale HAWT has played an important role in wind power generation in the rural and offshore areas. Hence, a large number of public opinion surveys on wind turbine noise are also focused on such turbine rotor. Although there is still no critical evidence that wind turbine noise has adverse health effects, the results from the quantitative, experimental and longitudinal research consistently suggest that wind turbine noise presents a risk for sleep and interference with work \cite{poulsen2019impact,karasmanaki2022safe}. Compared with the mechanical noise, the aerodynamic noise emitted from wind turbines is more significant, and it can be paralleled to the ‘swishing’ sound of a helicopter \cite{karasmanaki2022safe}. Since the low frequency wind turbine noise penetrating the building walls is more likely to cause indoor noise annoyance \cite{hongisto2017indoor}, there are stricter limits and requirements for wind turbine noise in urban areas. This also explains another reason why small VAWTs are more popular than HAWTs in urban areas, namely the former's less aerodynamic noise \cite{eriksson2008evaluation}.

\subsection{State of research on VAWT noise}
Considering that small VAWTs have more attractive features than HAWTs in urban areas, including lower cost, better adaptability to complex wind environments, and lower noise, researchers have begun to explore the noise characteristics of VAWTs and noise reduction methods. For instance, Botha et al. \cite{botha2017implementation} investigated the inflow turbulence noise and airfoil self-noise of a six-bladed 2kW VAWT by using the analytical flow methods and SST $k-\omega$ turbulence model. It was shown that the use of CFD calculations improved the accuracy of noise prediction. Besides, the inflow-turbulence noise was the major source, where blade generated turbulence dominated the atmospheric inflow turbulence. Aihara et al. \cite{aihara2021aeroacoustic} utilized the large eddy simulation and the Ffowcs Williams and Hawkings' (FW-H) acoustic analogy to solve the aerodynamic noise of a H-type VAWT. The results showed that the loading noise could be considered to be the main noise contributor, and a higher noise level was found in the downwind region. Besides, Liu et al. \cite{liu2022aerodynamic} combined the CFD technology and FW-H method to investigate the aerodynamic and aeroacoustic performance of straight-bladed wind turbine with active blowing-suction devices. The results indicated that the jet coefficient significantly determined the wind turbine performance, and the optimized active device could reduce the acoustic noise at low frequency. Recently, an experimental investigation on small VAWT noise in an urban environment was conducted by Li et al. \cite{li2022experimental}. The low-frequency noise was recorded and the impacts of wind turbine noise on humans and animals were analyzed.

\begin{figure}[htp]
	\centering
	\includegraphics[width=0.2\linewidth]{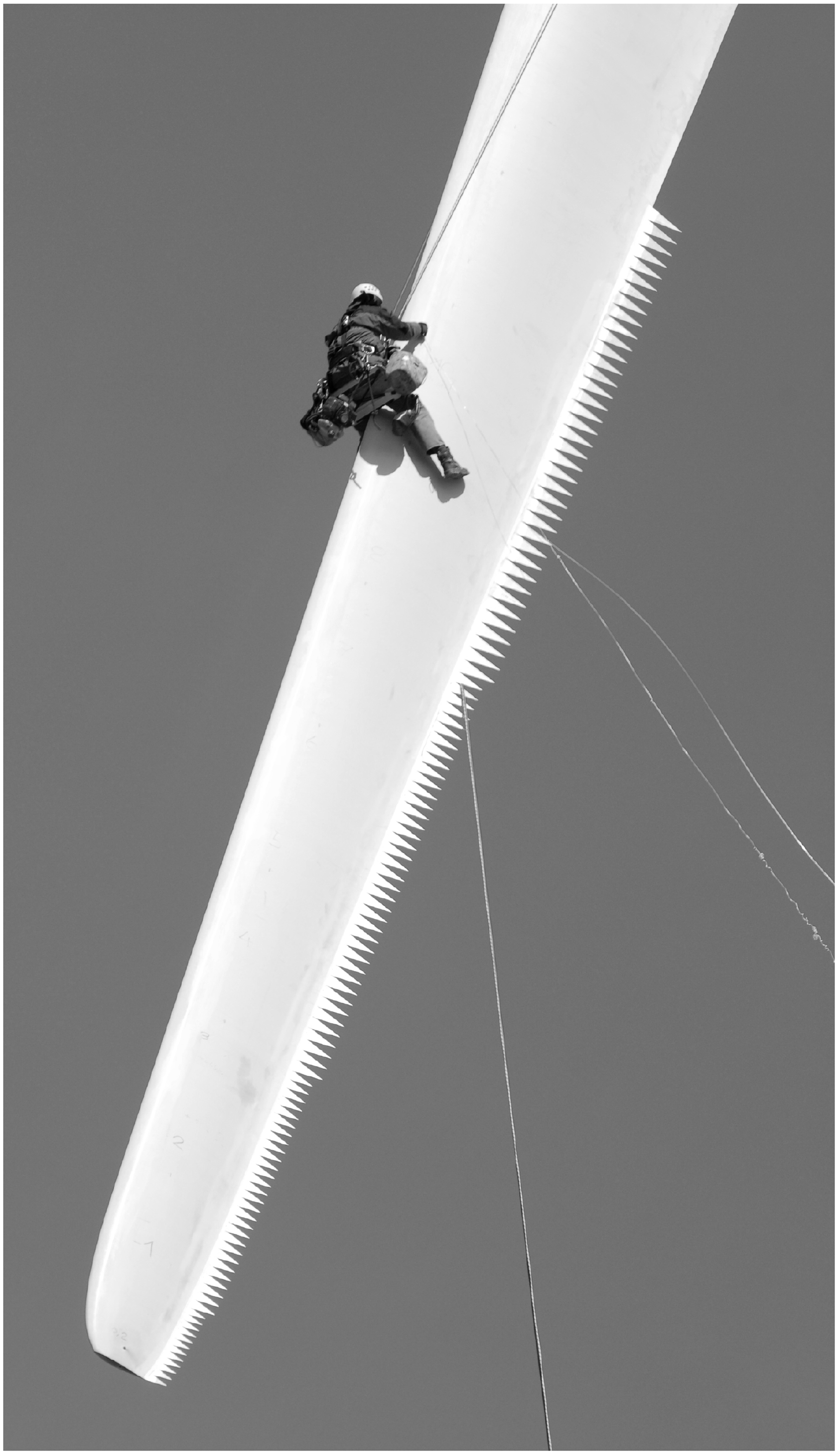}
	\caption{The serrated blade in a horizontal axis wind turbine \cite{oerlemans2009reduction}.}
	\label{fig:Fig1}
\end{figure}

For the aerodynamic noise reduction techniques, leading edge modification like tubercles, trailing-edge serrations or sawtooth structure, blade tip shape optimization, and boundary-layer trips, were introduced to reduce the pressure field fluctuation and the interaction between atmospheric turbulence and rotating blades. For the single wind turbine, Oerlemans et al. \cite{oerlemans2009reduction} investigated the aeroacoustics of a HAWT with different airfoils and trailing-edge serrations as shown in Fig. \ref{fig:Fig1}. The results showed that average overall noise reductions of 0.5 and 3.2 dB could be obtained for the optimized blade and the serrated blade. Maize et al. \cite{maizi2018noise} evaluated the effect of three tip blade configurations on the noise reduction of NREL Phase VI wind turbine by using CFD and FW-H methods. The shark tip blade was found to be the best tip configuration for optimal noise emission. Besides, Wu et al. \cite{wu2020optimizing} proposed an interesting strategy to minimize the noise of onshore wind farms without sacrificing power production. A stringent noise control strategy reducing noise by 11$\%$ and a flexible strategy reducing noise by 5.7$\%$ were reported.

On the other hand, there are relatively few studies on the application of these noise reduction techniques on VAWTs. Mohamed \cite{mohamed2016reduction} attempted to reduce the noise of a VAWT by using three couples of blades. It was reported that the 60$\%$ spacing was the best configuration of the double-airfoil for reducing wind turbine noise, while the power performance was not revealed. Last year, the stochastic noise from a model-scale VAWT was tested by Graham and Pearson \cite{graham2022noise} in a wind tunnel in the UK. It was found that the surface trips effectively reduced the laminar-instability noise, and the influence of relative blade speed on noise is greater than the number of blades.

\subsection{Motivation and contribution of this work}
Through the above literature review, it can be found that considerable research efforts have been devoted to improving the power output of VAWTs, but there is still more space for improvement in the wind turbine structure. On the other hand, while a number of research on wind turbine noise have been carried out on, there are few studies on the aerodynamic noise of VAWT other than straight-bladed wind turbine. The directivity and propagation of such wind turbine noise are not very clear. Besides, the problem of noise pollution prompts the VAWT to reduce noise while increasing power performance. The possibility of applying noise reduction techniques on VAWTs remains to be explored. 

Therefore, the present study attempts to make further investigation on the aerodynamic and aeroacoustic performance of the small VAWT with V-shaped blades and trailing-edge serrations based on the authors’ previous study \cite{su2020investigation}. In this study, a series of V-shaped blades were investigated to evaluate their effect on the aerodynamic and aeroacoustic characteristics of the VAWT. The unsteady Reynolds-Averaged Navier-Stokes (URANS) SST \textit{k}-$\omega$ turbulence model was utilized to perform the simulations, and the computational settings were demonstrated in Section 2. Then a brief description of FW-H analogy used for acoustic calculation was presented in Section 3. In Section 4, a series of three-dimensional CFD simulations were carried out to evaluate the effects of different V-shaped blades on the wind turbine performance at different tip speed ratios. Besides, the components of tonal noise were discussed, as well as the results of noise directivity and propagation. Moreover, the trailing-edge serrations were employed in the VAWT to evaluate the noise reduction effect. Finally, several conclusions were summarized in Section 5.

The contributions of this study are summarized as follows:

(i) The present study comprehensively evaluates the effects of a series of V-shaped blades on the aerodynamic forces of a small VAWT at different tip speed ratios. The improved blades can effectively increase wind turbine power output over a fairly wide range of tip speed ratios while only slightly increasing the thrust loads on the structure.

(ii) This study verifies the elliptical distribution of noise in the VAWT with straight blades and a dumbbell-shaped noise directivity distribution is discovered for the first time in VAWT, which was previously only found in HAWTs. Besides, the V-bladed VAWT generates less low-frequency noise. It can be explored for optimizing wind farms to minimize noise disturbance to potential receivers.

(iii) The trailing-edge serrations that can be manufactured separately and then installed on the wind turbine blade were successfully applied in the simulation of VAWTs. The expected noise reduction effect was achieved, and even a small increase in power output was realized. The combination of V-shaped blades and trailing-edge serrations can be practically applied to the design of high-efficiency and low-noise wind turbines.

\section{Methodology}

\subsection{Wind turbine model}

\begin{figure}[htbp]
	\centering
	\subfigure{
		\begin{minipage}[t]{1\linewidth}
			\centering
			\includegraphics[width=0.7\linewidth]{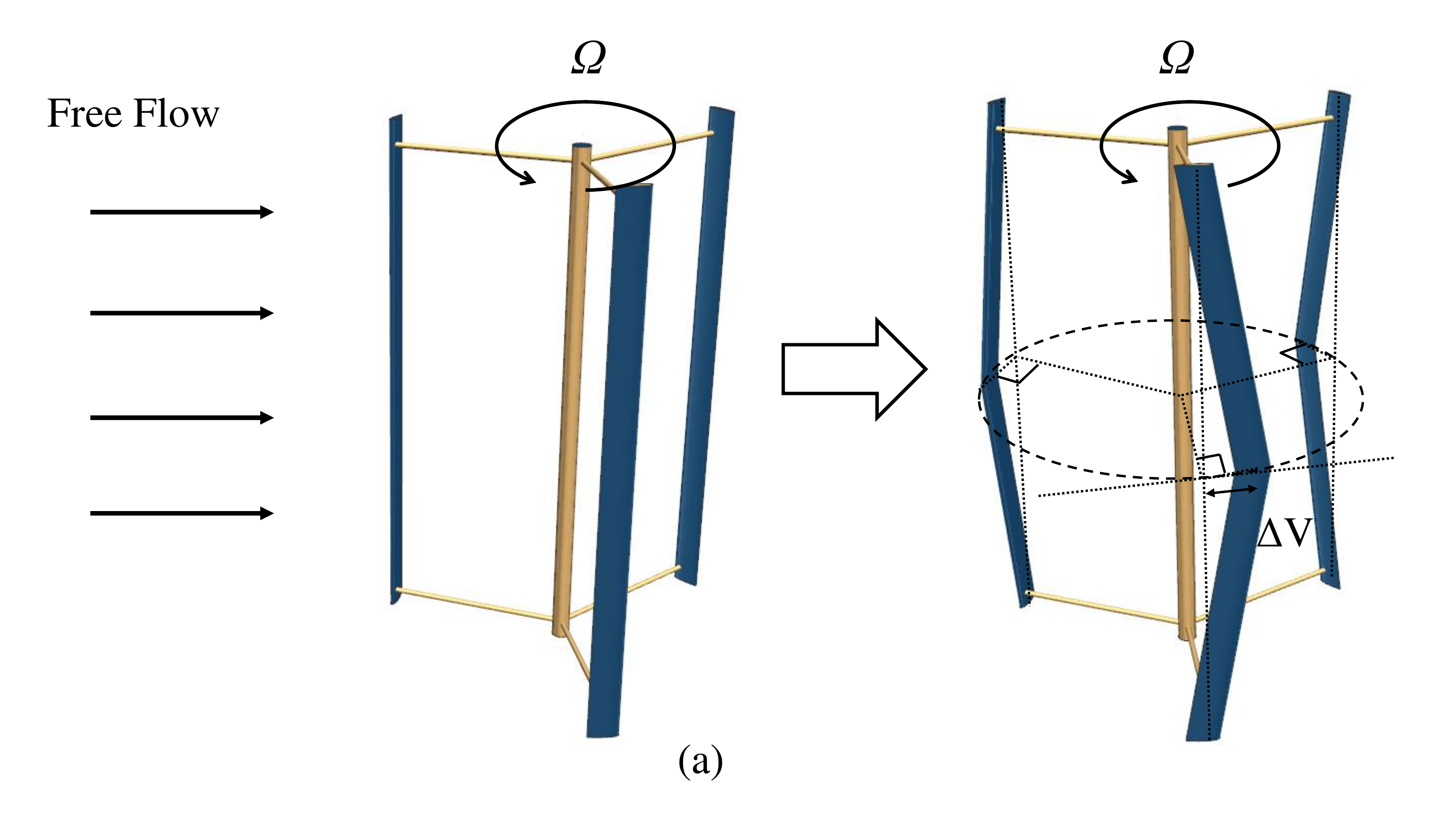}	
	\end{minipage}}
	
	\subfigure{
		\begin{minipage}[t]{1\linewidth}
			\centering
			\includegraphics[width=0.7\linewidth]{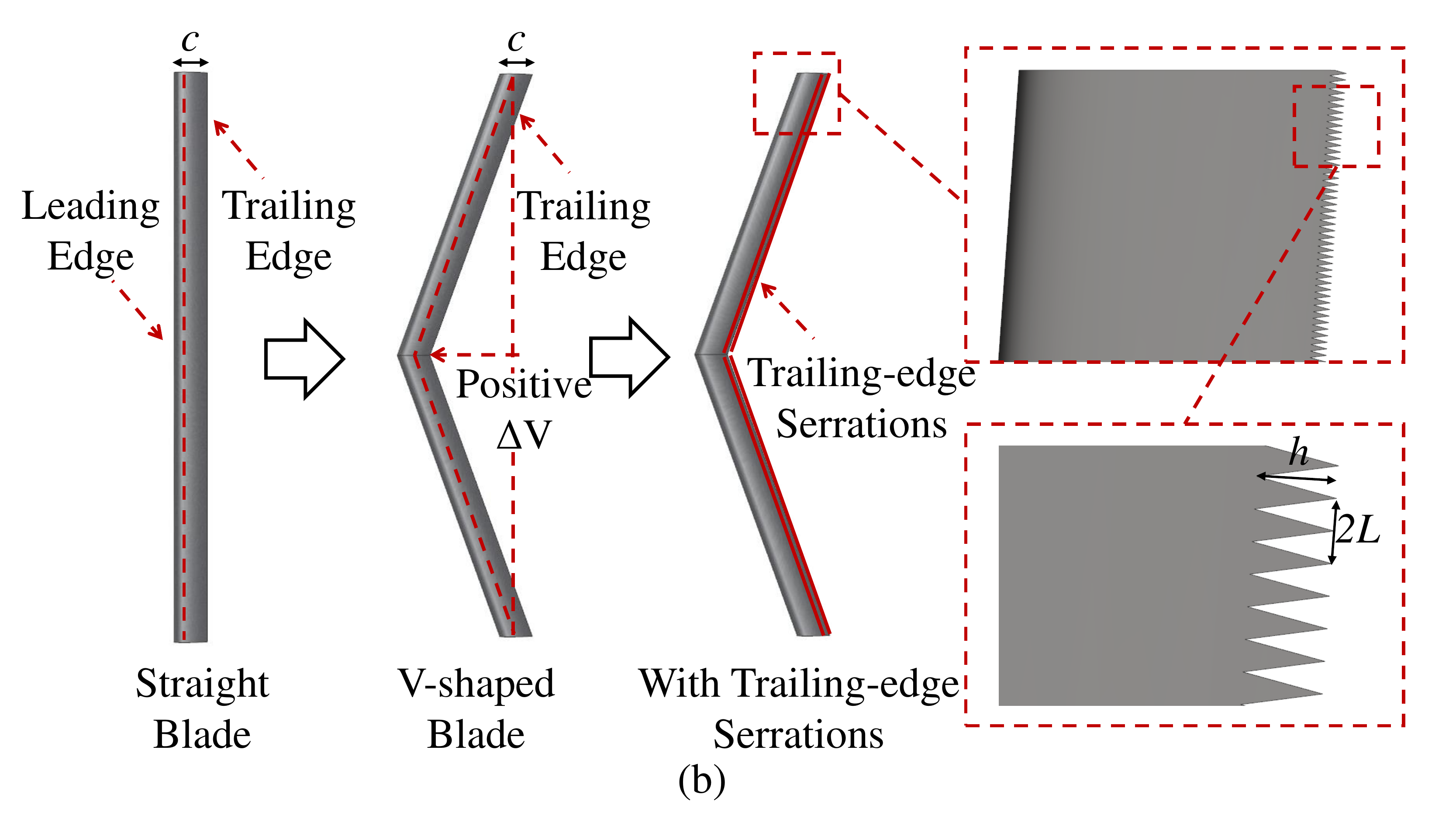}
	\end{minipage}}
	\caption{The sketch map of the V-bladed VAWT: (a) transformation of the wind turbine model; (b) the structures of V-shaped blade and trailing-edge serrations.  }
	\label{fig:Fig2} 
\end{figure}

The selected VAWT with V-shaped blades was modified from the prototype of the straight-bladed wind turbine tested by Castelli et al.\cite{castelli2010modeling}. The airfoil NACA0021 was utilized to produce the wind turbine blades both in the V-bladed and straight-bladed VAWTs. The geometrical characteristics of the reference wind turbine, as well as the operating conditions are listed in Table \ref{table 1}. The schematics of the two types of wind turbines are illustrated in Fig. \ref{fig:Fig2} (a). In accordance with the previous study \cite{su2020investigation}, the V-shaped blade was created by moving the position of the blade middle cross section along the tangential direction of the rotor. The structure of the V-shaped blade is controlled by the distance between the leading edge of the modified blade and that of the straight blade, which can be defined as $\Delta V$. In this study, a series of V-shaped blades would be investigated to evaluate their aerodynamic performance and noise generation. The more detailed construction method of the V-shaped blade is presented in Ref.\cite{su2020investigation} and for brevity is not repeated here. Besides, for the sake of computational cost, the shafts and struts are neglected.

\begin{table}[h]
	\caption{Geometric properties of the VAWT model.}
	\centering
	\begin{tabular}{l c l}
		\toprule
		Property& Symbol& Value\\
		\hline
		Airfoil profile& - & NACA 0021\\
		Blade number&$N$& 3\\
		Chord length&$c$& 0.0858 m\\
		Rotor diameter&$D$&1.030 m\\
		Span length&$H$&1.456 m\\
		Swept area of rotor&$A$&1.236 m$^{2}$\\
		Solidity&$\sigma$&0.25\\
		Tip speed ratio&$\lambda$& 1.5 - 3.3\\
		\bottomrule
	\end{tabular}
	\label{table 1}
\end{table}

On the other hand, to further explore the noise reduction effect, the trailing-edge serrations were applied to the wind turbine blade as shown in Fig. \ref{fig:Fig2}(b). Among the noise reduction techniques mentioned in Introduction, the trailing-edge serrations were considered one of the relatively practical methods. Given that the serrated structures can either be processed directly on the blade, or can be fabricated separately and then installed on the blade, this structure has great flexibility in processing and installation. It can be used as an optional component for wind turbine blades as shown in Fig. \ref{fig:Fig1}. In the present study, the trailing-edge serrations were created by directly processing the trailing edge of the V-shaped blade as illustrated in Fig. \ref{fig:Fig2}(b). The serrated structure can be defined by the wavelength $L$ and root-to-tip amplitude $2h$. In this preliminary study, the serrated structure with $L/h=0.4$ and $h/c=0.05$ was adopted as suggested by Lyu et al. \cite{lyu2016prediction}, which was fully distributed along the blade span.

The power coefficient $C_{P}$, torque coefficient $C_{Q}$, and thrust coefficient $C_{thrust}$ of the wind turbine can be defined as:
\begin{align}
	C_{P}=\frac{Q\Omega}{\frac{1}{2}\rho U_{\infty}^{3}A}\\
	C_{Q}=\frac{Q}{\frac{1}{2}\rho U_{\infty}^{2}AR}\\
	C_{thrust}=\frac{F_{thrust}}{\frac{1}{2}\rho U_{\infty}^{2}A}
\end{align}
where $Q$, $\Omega$, $A$, $R$, $\rho$, and $U_{\infty}$ are the rotor torque, angular velocity, swept area of the rotor, rotor radius, air density, and freestream velocity, respectively.

\subsection{Computational settings}

\begin{figure}[htp]
	\centering
	\includegraphics[width=1\linewidth]{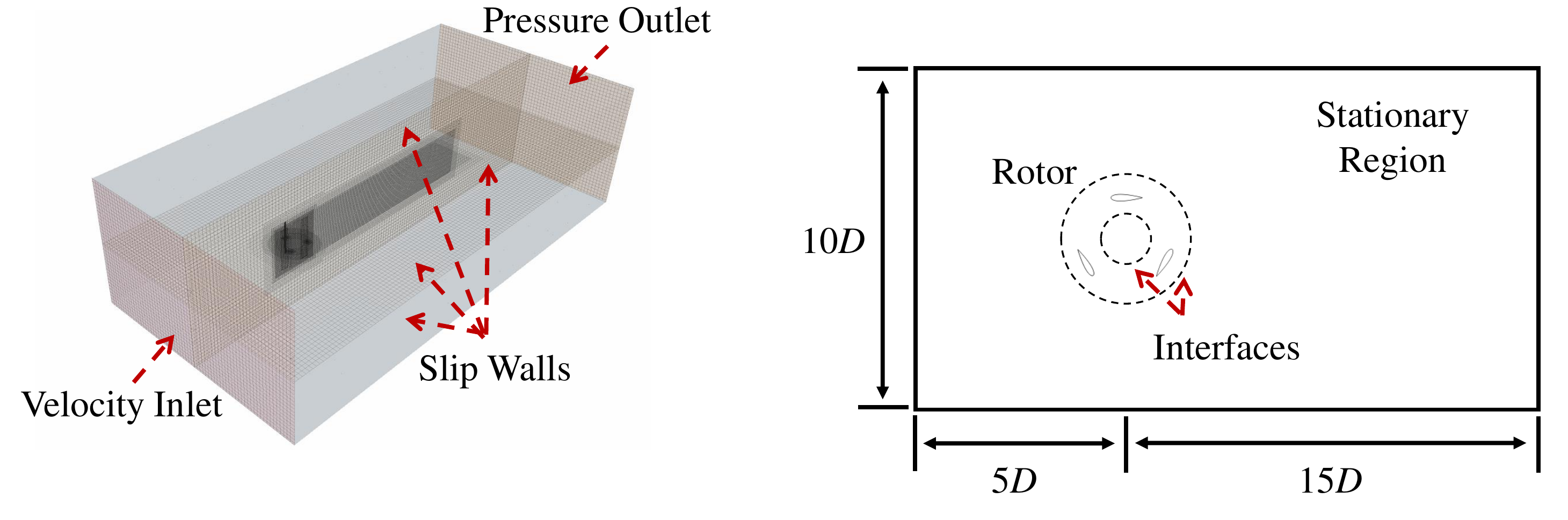}
	\caption{The three-dimensional computational model and boundary conditions.}
	\label{fig:Fig3} 
\end{figure}

The computational model and mesh topology are shown in Fig. \ref{fig:Fig3}. The size of the computational model was $20D \times 10D\times4H$. The distance between the outlet boundary and the VAWT was set as 15$D$ to ensure the full development of wind turbine wake. The computational model was divided into a stationary domain and a rotational domain, where the latter was used to simulate the rotational motion of the wind turbine by utilizing the sliding mesh technique. Besides, two refined regions were set up to obtain the accurate flow field and the pressure fluctuation information which would be used in the acoustic calculation. The four sides of the computational model were set as slip walls, while the inlet and the outlet boundary were set as uniform freestream velocity inlet ($U_{\infty}$=9 m/s) and pressure outlet (0.0 Pa). Moreover, the density of the air and the turbulence intensity were set as $\rho=1.225$ kg/m$^{3}$ and 5$\%$ to reproduce the experimental conditions as much as possible. 

\begin{figure}[htp]
	\centering
	\includegraphics[width=0.5\linewidth]{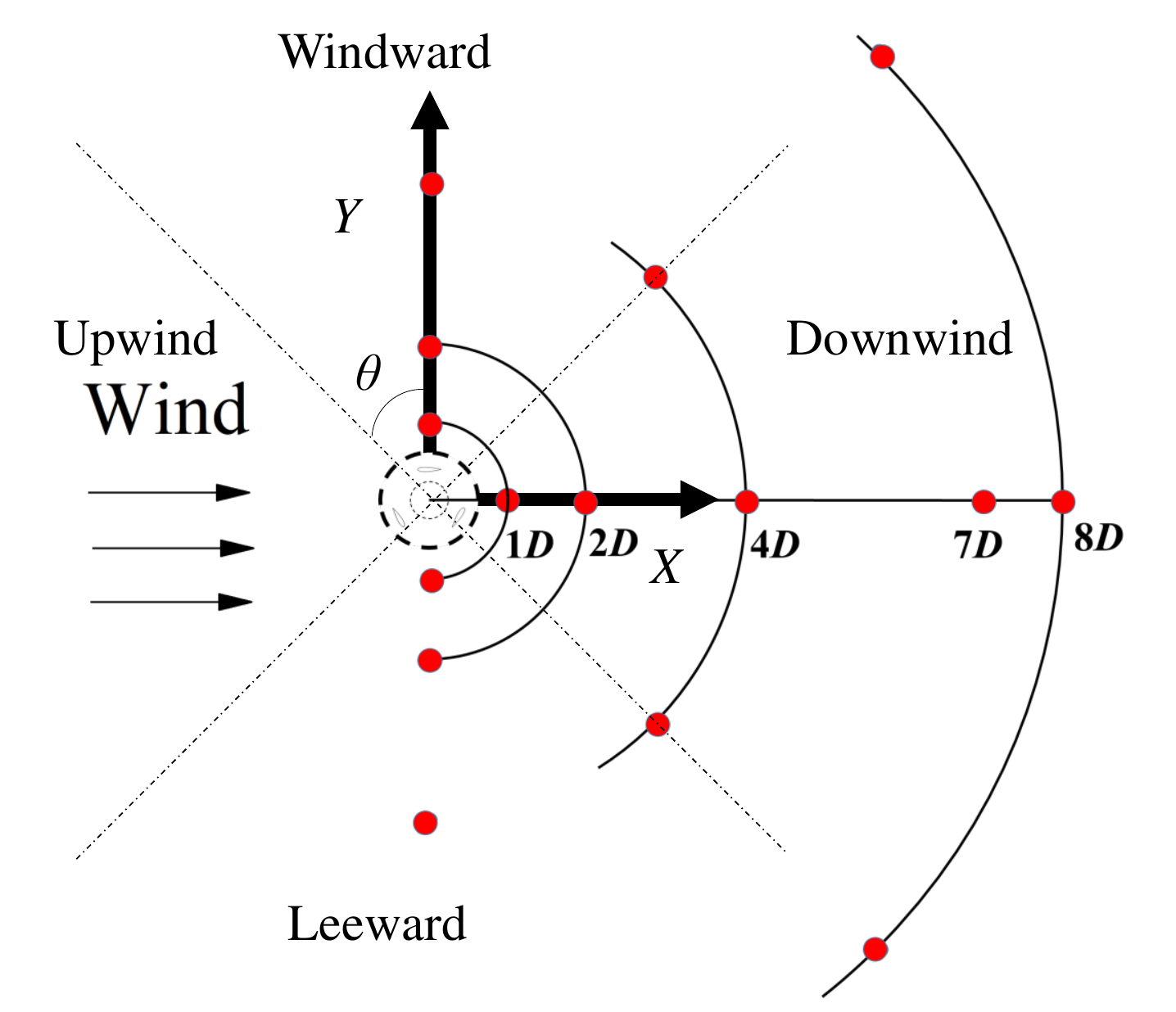}
	\caption{The distribution of probes for acoustic calculation.}
	\label{fig:Fig4} 
\end{figure}

In order to facilitate the analysis of the results, four regions are defined in the model based on the azimuthal range of the blade. The region ranged from $45^\circ<\theta\leq135^\circ$, $135^\circ<\theta\leq225^\circ$, $225^\circ<\theta\leq315^\circ$, and $315^\circ<\theta\leq405^\circ$ ($45^\circ$) are defined as the upwind region, leeward region, downwind region and windward region, respectively. Apart from that, a total of 33 probes for acoustic calculations were set up in the middle plane of the computational domain as illustrated in Fig.\ref{fig:Fig4}. The position distribution of the other 66 probes in the bottom and top planes near the blade tips is the same as that in the middle plane. The position statistics of the probes are shown in Appendix A.

The numerical settings employed in the present study are based on the guidelines for high-fidelity simulations of VAWTs \cite{rezaeiha2018towards}, and these settings were successfully applied in the previous study \cite{su2020investigation}. In this study, the software STAR-CCM+ was utilized to simulate the three-dimensional incompressible unsteady flow around the VAWT. The SST $k-\omega$ turbulence model was applied in the current CFD simulations. The SST $k-\omega$ model is feasible to obtain the flow field information around the wind turbines, and the acoustic calculation based on this model is also credible \cite{mohamed2019criticism,sobhani2017numerical,tadamasa2011numerical}. The convection terms and the temporal discretization were set as second order. The SIMPLE algorithm was applied for coupling the pressure-velocity equation. The time-step size was set as $2^\circ$ \cite{su2019aerodynamic,lei2017three} in the first eight revolutions to obtain the instantaneous flow field, and 20 iterations were conducted in each time step to ensure that the residuals are small enough. After that, the aeroacoustic calculation were performed in another two revolutions with FW-H method which would be presented latter. In the acoustics calculation, considering the convective Courant number and the sensitive frequency band of human hearing, the time-step size was set as $2\times10^{-5}$s. The Reynolds number calculated in this study is about  $Re$ = 1.1$\times10^{5}\sim2.2\times10^{5}$. All simulations were carried out on the small-scale Server with two Intel(R) Exon(R) CPUs (E5-2673 v3, 2.40 GHz). About 168 h were taken to calculate the eight revolutions and about 360 h were used to accomplish the acoustic calculation.

\subsection{Validation studies}
\begin{figure}[htbp]
	\centering
	\includegraphics[width=0.5\linewidth]{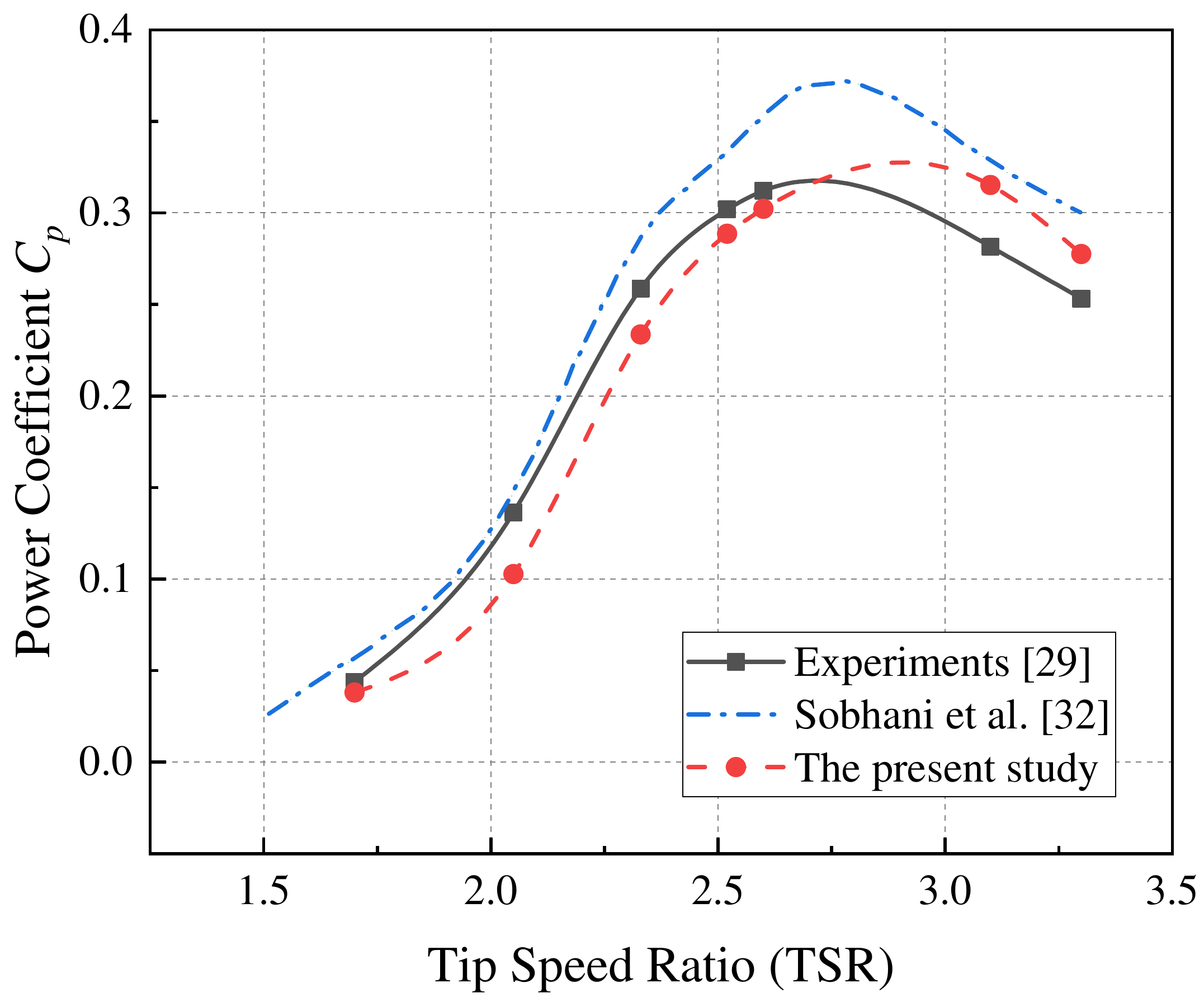}
	\caption{Comparison of the current results against experimental and numerical data \cite{castelli2010modeling,sobhani2017numerical}.   }
	\label{fig:Fig5} 
\end{figure}

A set of validation studies have been conducted on the selected VAWT, including the grid independence analysis and comparison of target functions $C_{P}$. A good agreement between the present CFD results and the experiments can be observed as shown in Fig. \ref{fig:Fig5}. The results of the medium mesh scheme with a total of 11.7 million grids are found to be sufficiently independent of mesh resolution, where the wall function $y^+$ was approximately equal to 1 during the calculation. The present study is conducted based on the authors’ previous research, and the more detailed description of the validation tests can be found in Ref. \cite{su2020investigation}, which will not be repeated here.

In addition, another grid sensitivity analysis of serrated V-bladed VAWT was carried out. Table \ref{table 2} shows the aerodynamic performance with different mesh refinement levels. The results show that the average deviation of torque is small enough when the mesh is refined from TE Refinement II to TE Refinement III, resulting in less than 1$\%$ errors in power coefficient. It can be observed that the results of mesh scheme TE Refinement II are sufficiently independent of mesh resolution.   

\begin{table}[h]
	\caption{Mesh topology around the serrated blade at $\lambda=2.6$.}
	\centering
	\begin{tabular}{l l l l}
		\toprule
		Mesh scheme& Minimum Grid Size&Total Number of Grids&  $C_{P}$\\
		\hline
		TE Refinement I&12$\times10^{-4}$ m & 21.7 million & 0.3612\\
		TE Refinement II&6$\times10^{-4}$ m & 35.4 million & 0.3877\\
		TE Refinement III  &3$\times10^{-4}$ m & 54.8 million & 0.3904\\
		\bottomrule
	\end{tabular}
	\label{table 2}
\end{table}

Overall, the present model is proved to successfully replicate the experimental results, and the simulations can reflect the aerodynamic of real-life wind turbine in wind tunnel. Therefore, it suggests that the present numerical model can be applied as a credible approach in following studies.

\section{Acoustic calculation}
In this study, the Ffowcs Williams and Hawkings’ acoustic analogy method is utilized to predict the wind turbine noise generated from the straight-bladed and V-bladed VAWTs. To obtain the acoustic pressure fluctuations at each receiver as described above, the time-accurate solutions of the flow-field variables around the VAWT are required from the SST $k-\omega$ model \cite{mohamed2019criticism,tadamasa2011numerical}. 

In order to assess the strength of the wind turbine noise and perform a comprehensive analysis on the directivity of noise propagation, 33 acoustic receivers were set up in the middle plane to cover the sound wave propagation. The details about the receivers can be found in Section 2.2.

Besides, to enhance the credibility of calculation results obtained by using the SST $k-\omega$ model and the FW-H method, taking into account the incomplete acoustic experimental data of the selected VAWT, another verification test of sound generated by flow around a circular cylinder is shown in Appendix B.


\section{Results and discussions}

In this section, a series of V-shaped blades were investigated to evaluate their impact on the wind turbine aerodynamic performance and acoustic characteristics. The aerodynamic forces and noise spectrum were analyzed, and the effect of trailing-edge serrations was evaluated.

\subsection{Aerodynamic performance assessment}

\begin{figure}[htbp]
	\centering
	\subfigure[]{
		\begin{minipage}[t]{0.485\linewidth}
			\centering
			\includegraphics[width=1\linewidth]{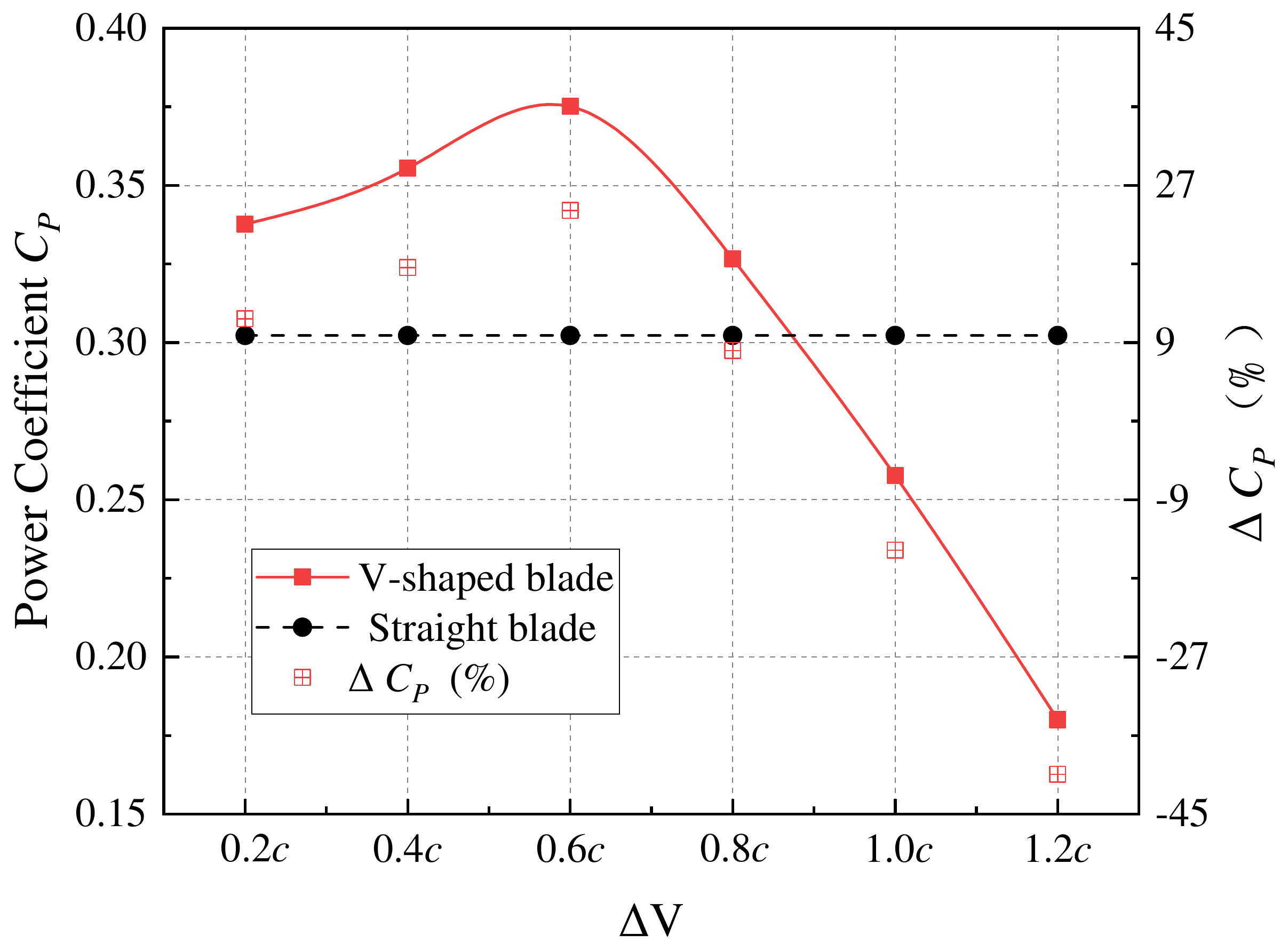}	
	\end{minipage}}	
	\subfigure[]{
		\begin{minipage}[t]{0.485\linewidth}
			\centering
			\includegraphics[width=1\linewidth]{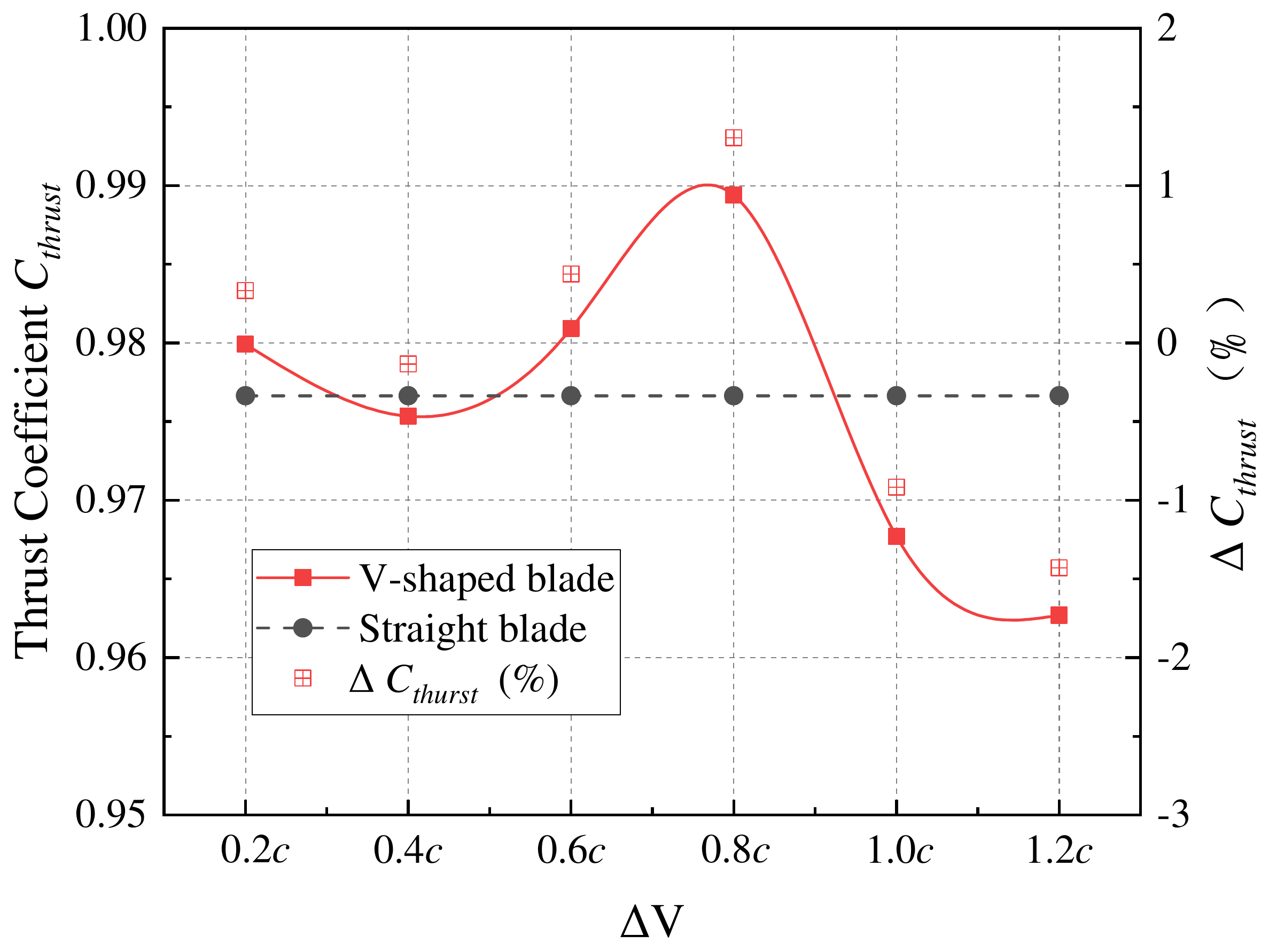}
	\end{minipage}}
	
	\caption{Comparison of the VAWT with different V-shaped blades at the optimal tip speed ratio $\lambda=2.6$: (a) power coefficient; (b) thrust coefficient.   }
	\label{fig:Fig6} 
\end{figure}

The power coefficient and thrust coefficient are the key parameters to assess the wind turbine’s energy conversion efficiency and the load characteristics. Fig. \ref{fig:Fig6} presents the comparison of average power coefficient and thrust coefficient for the VAWT with different V-shaped blades at the optimal tip speed ratio $\lambda=2.6$. It can be observed that the modification of the original straight blade would greatly affect the torque output of the wind turbine. The power coefficient of the VAWT gradually increases with the value of $\Delta V$, and the maximum enhancement of $C_{P}$ occurs at $\Delta V=0.6c$. According to the previous published study, most positive power outputs are attributed to the middle segments of the blade, while the effect of V-shaped structure on the blade tip region is relatively small. After that, increasing the value of $\Delta V$ has a less positive effect on the power performance of the wind turbine, and even larger $\Delta V$ may lead to performance degradation. On the other hand, the influence of V-shaped blade on the thrust coefficient of the wind turbine is relatively small, and the variation of $C_{thrust}$ is less than 4$\%$ in the studied cases.

\begin{figure}[htbp]
	\centering
	\subfigure[]{
		\begin{minipage}[t]{0.485\linewidth}
			\centering
			\includegraphics[width=1\linewidth]{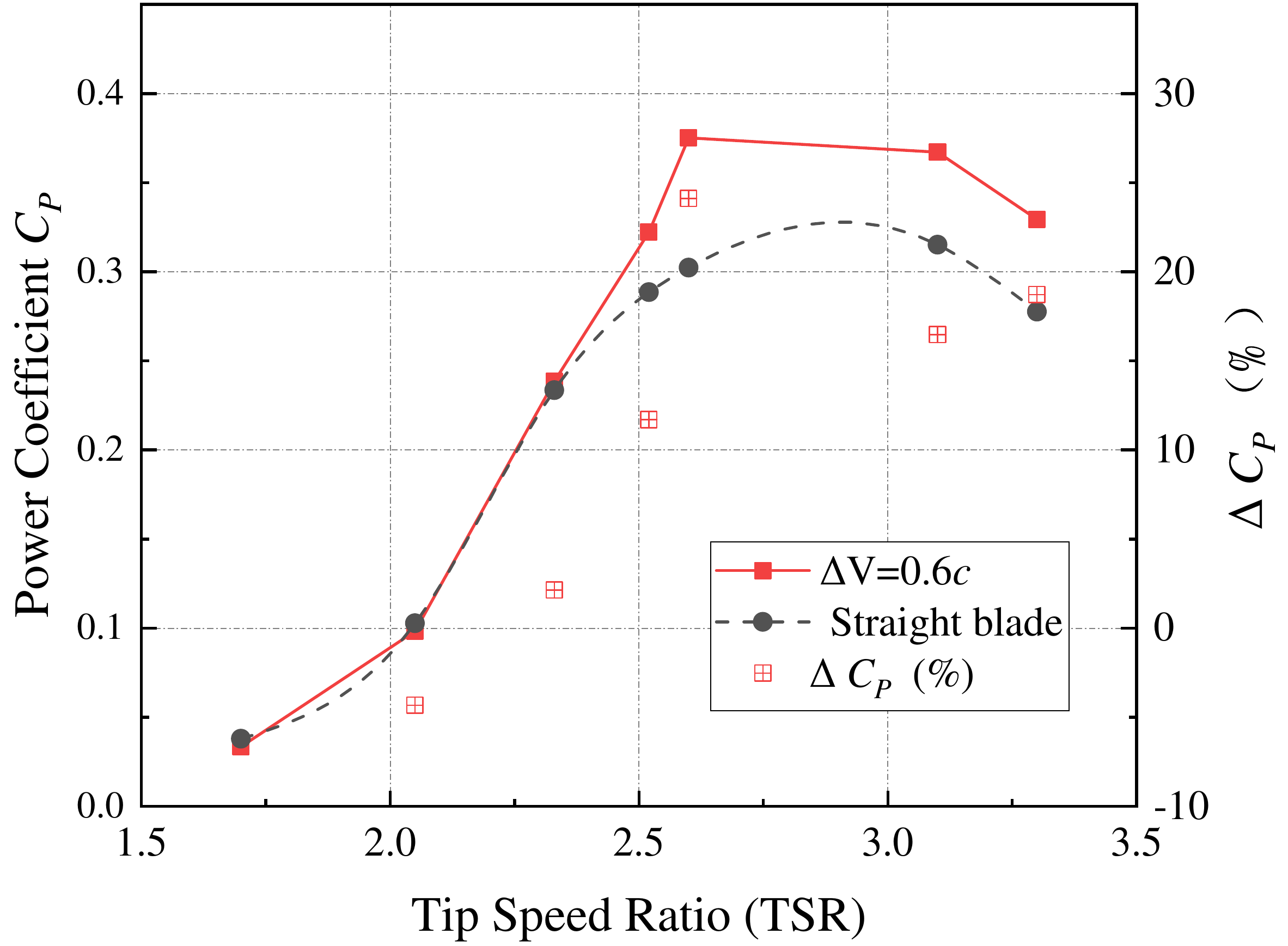}	
	\end{minipage}}	
	\subfigure[]{
		\begin{minipage}[t]{0.485\linewidth}
			\centering
			\includegraphics[width=1\linewidth]{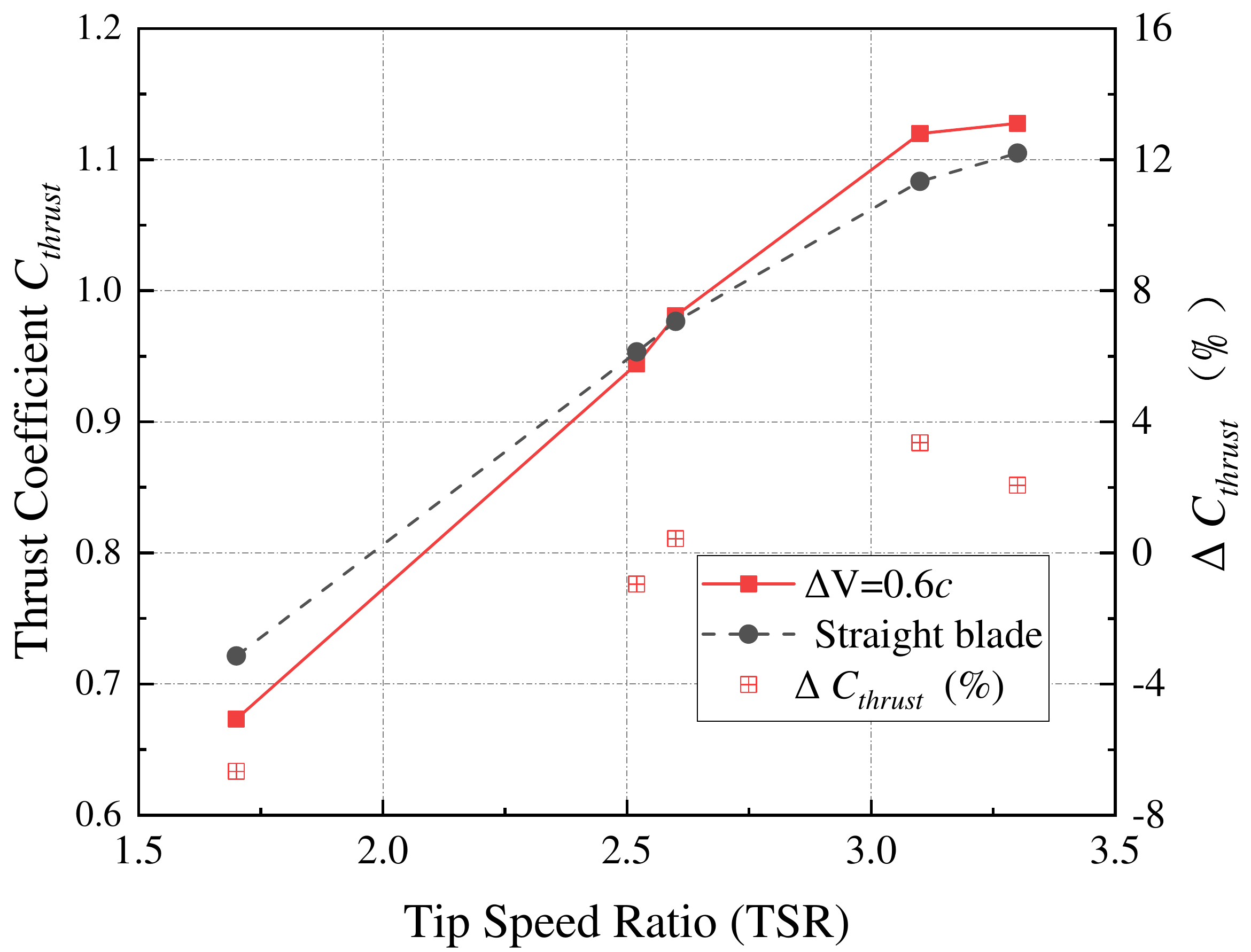}
	\end{minipage}}
	
	\caption{Comparison of the power coefficient  and thrust coefficient under different tip speed ratios between V-shaped blade and  straight blade: (a) power coefficient; (b) thrust coefficient.
	  }
	\label{fig:Fig7} 
\end{figure}

Based on previous results, the modified blade $\Delta V=0.6c$ with the best performance in power coefficient was selected and studied at different tip speed ratios. As shown in Fig. \ref{fig:Fig7} (a), it is clear that the V-shaped blade with $\Delta V=0.6c$ improves the power outputs from moderate to high tip speed ratios. In the previous study, it has been demonstrated that the V-shaped blades effectively increase the torque output in a small part of the upwind region and half of the leeward region ($115^\circ<\theta\leq180^\circ$), where the drag coefficient of blade is significantly reduced and the local mutation of drag coefficient would be eliminated. Besides, for the whole rotor, the fluctuation of torque output of the V-bladed VAWT is smaller than that of baseline turbine (see Fig. \ref{fig:Fig8}). The more detailed explanation can be found in Refs. \cite{su2020investigation,zhang2020laminar}. Over all tip speed ratios, the improved blade increases the maximum power coefficient by 24.1$\%$, while the value of $C_{P}$ is reduced at low tip speed ratios. The statistical results show that the average power increment for the modified blade from moderate to high tip speed ratios is 14.63$\%$.

Similarly, the thrust coefficient was investigated under all operating conditions. It is well known that the thrust coefficient of the wind turbine increases with the rotational speed. Compared with the variation curve of thrust coefficient of the straight-bladed VAWT, it can be found that the V-shaped blades do not alter this trend. However, it is noticed that the thrust coefficient of the modified wind turbine blade is less than that of the original one at low tip speed ratios, while the value of $C_{thrust}$ is higher than the baseline value at high rotational speed. Although the increment in $C_{thrust}$ is relatively small at high tip speed ratios, in terms of structural design, more consideration should also be given to the loads on the wind turbine blades and tower.

\subsection{Acoustic characteristics}
\begin{figure}[htbp]
	\centering
	\includegraphics[width=0.6\linewidth]{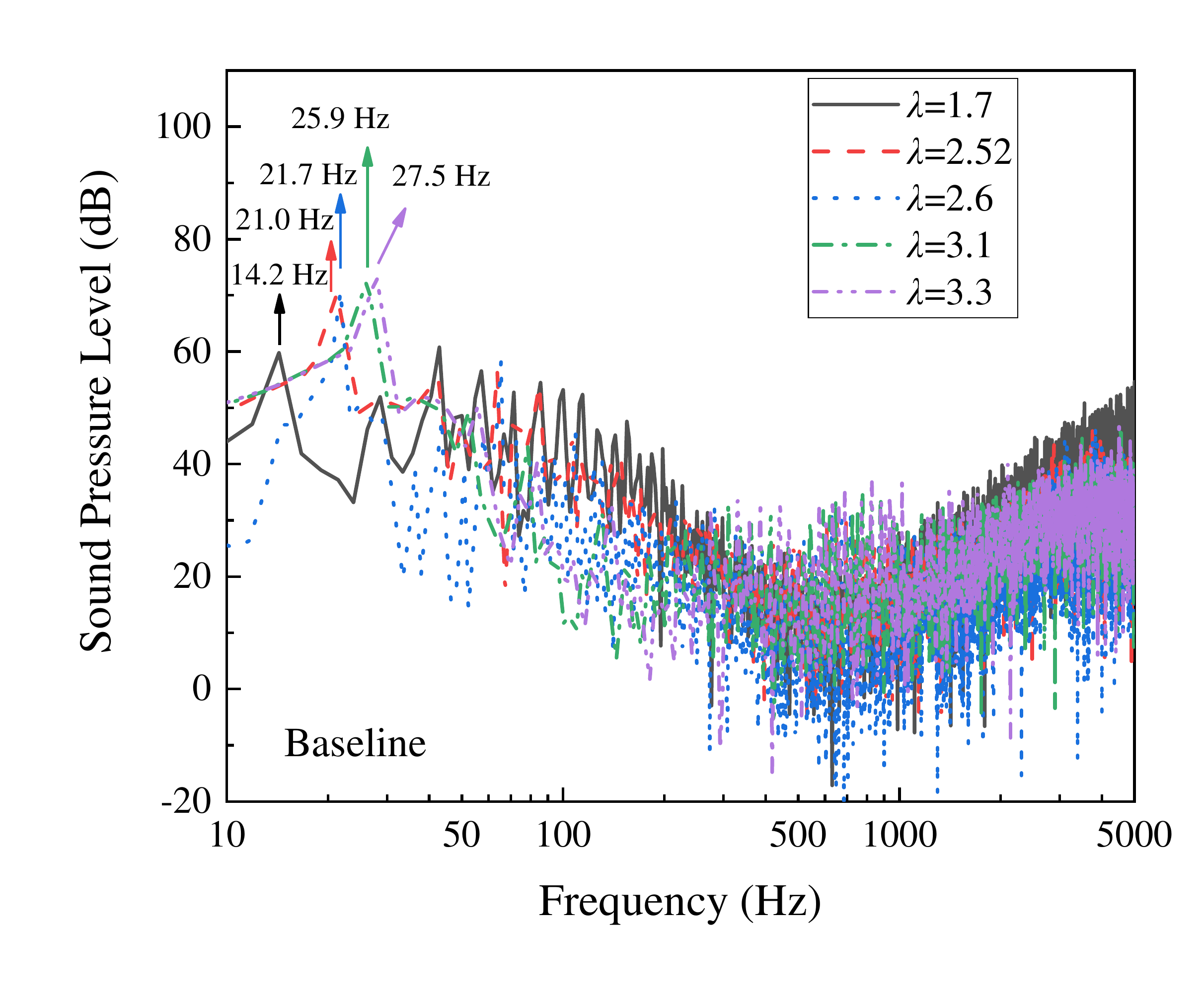}
	\caption{The sound pressure level of the baseline VAWT under different tip speed ratios.}
	\label{fig:Fig9} 
\end{figure}

In the following sections, only the baseline model and the V-shaped blade $\Delta V=0.6c$ with the best performance in $C_{P}$ are investigated at different tip speed ratios for acoustic analysis. To study the acoustic characteristics of VAWTs, Fig. \ref{fig:Fig9} presents the noise spectra of studied VAWT at different tip speed ratios. It can be found that the distinguishing tonal peaks appear at $f$=14.2, 21.0, 21.7, 25.9, and 27.5 Hz corresponding to the rotational speeds of wind turbine blade. Besides, the higher tip speed ratio usually induces larger noise level especially at low frequency. This trend is not changed in the acoustic spectrum of the VAWT with modified blades.

\begin{figure}[htbp]
	\centering
	\subfigure[]{
		\begin{minipage}[t]{0.48\linewidth}
			\centering
			\includegraphics[width=1\linewidth]{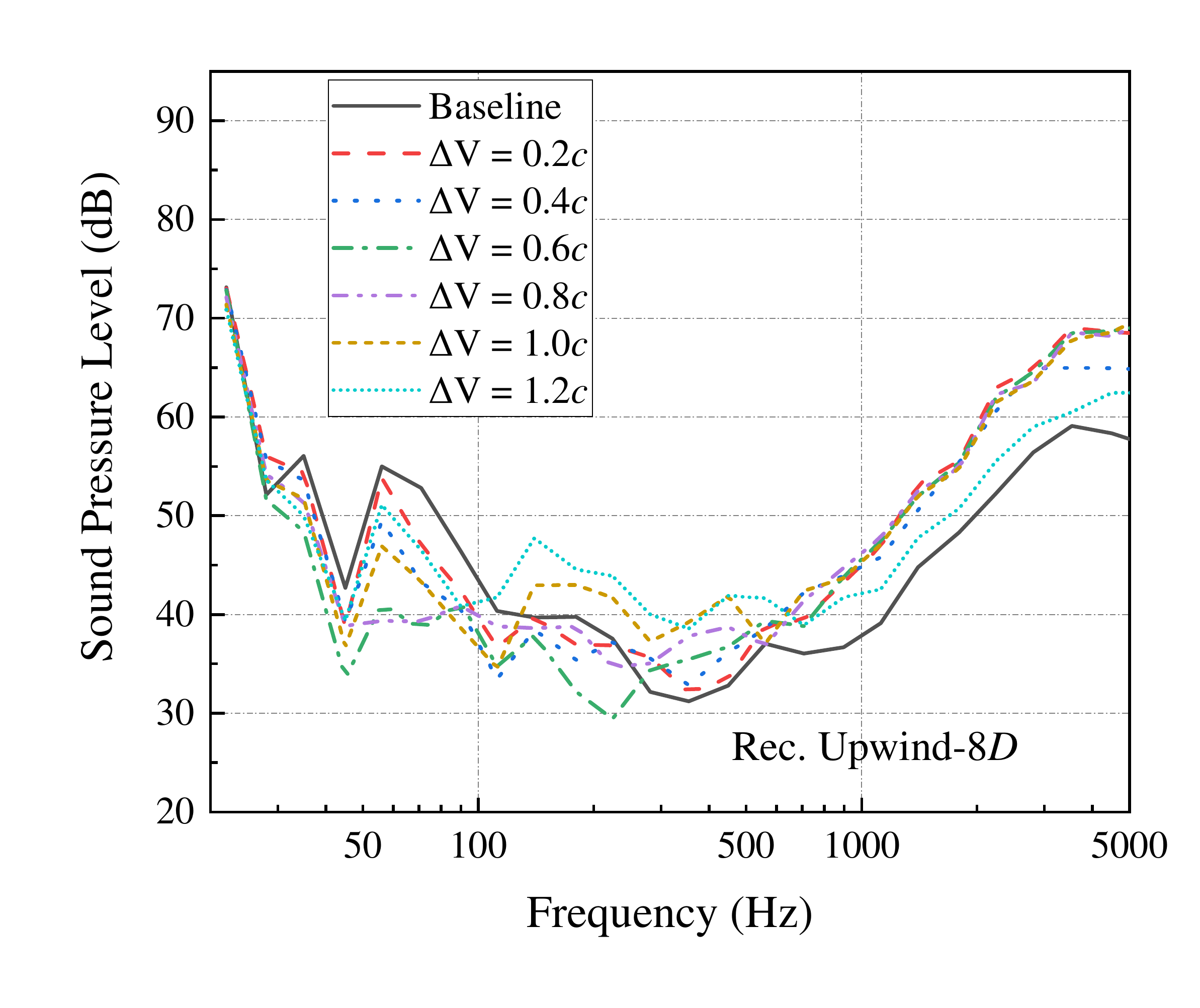}	
	\end{minipage}}	
	\subfigure[]{
		\begin{minipage}[t]{0.48\linewidth}
			\centering
			\includegraphics[width=1\linewidth]{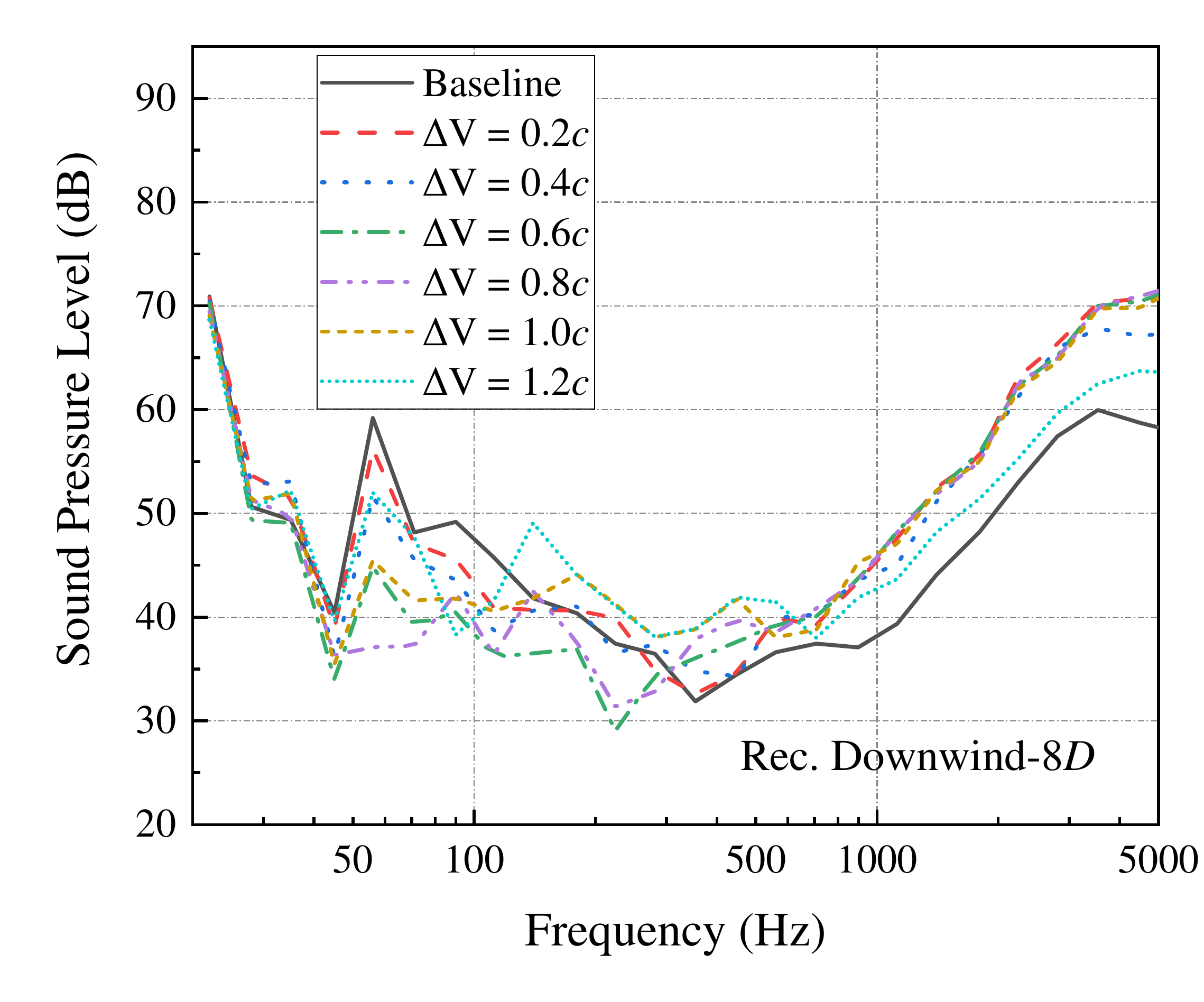}
	\end{minipage}}
	
	\subfigure[]{
		\begin{minipage}[t]{0.48\linewidth}
			\centering
			\includegraphics[width=1\linewidth]{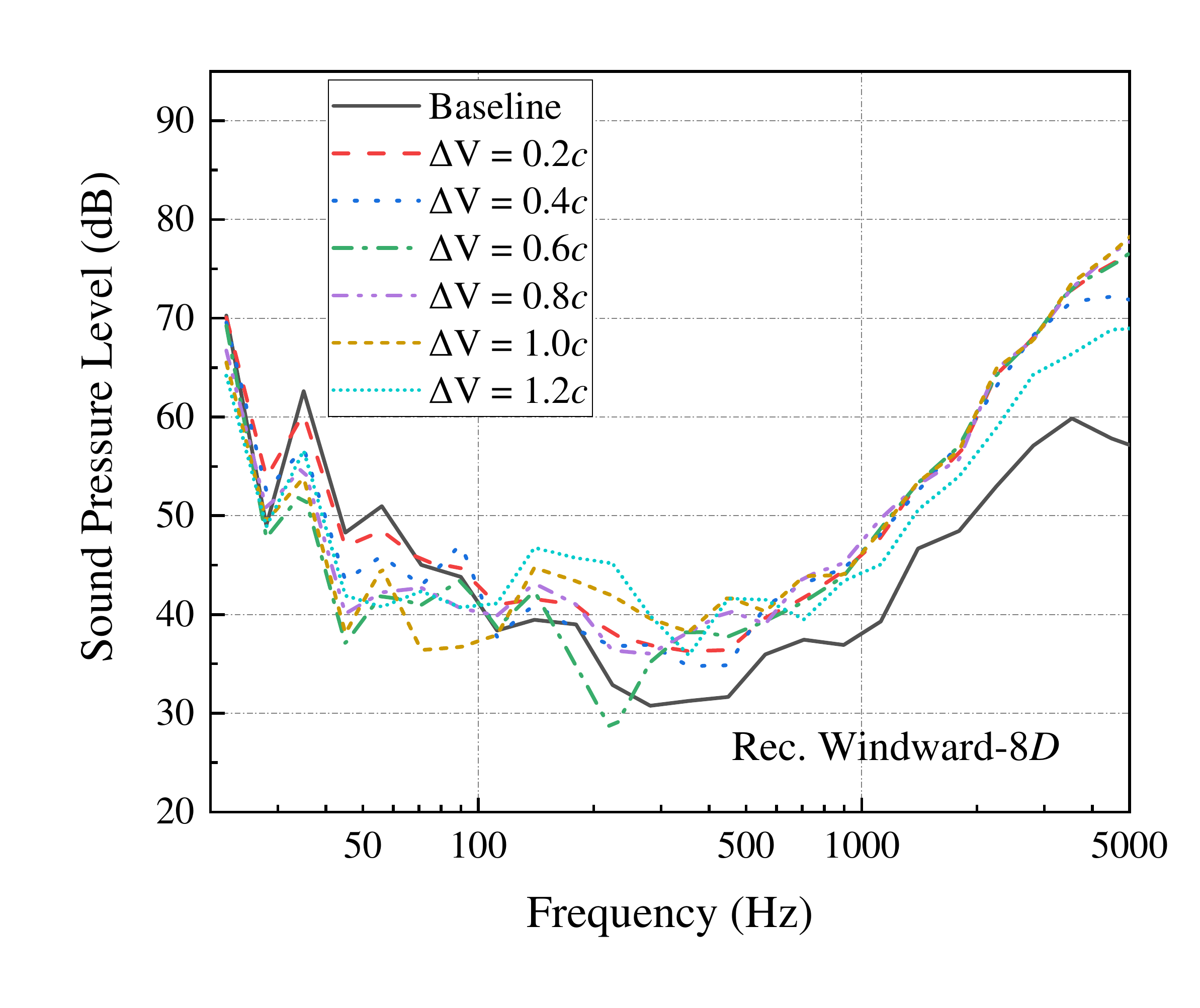}
	\end{minipage}}
	\subfigure[]{
		\begin{minipage}[t]{0.48\linewidth}
			\centering
			\includegraphics[width=1\linewidth]{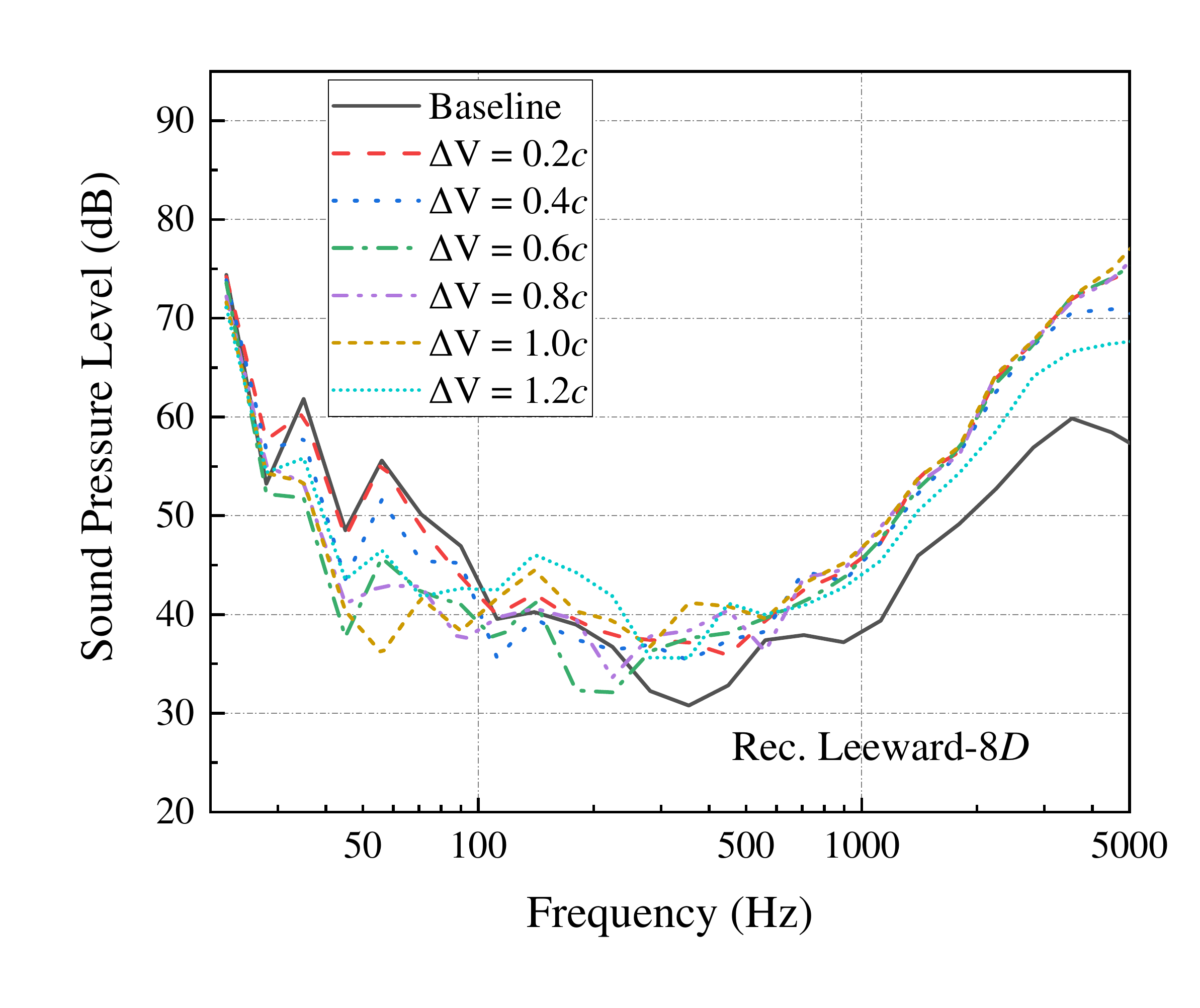}
	\end{minipage}}
	\caption{The 1/3 octave band spectra of the sound pressure level observed at a radial distance of 8$D$ from the center of rotor in four directions at $\lambda=2.6$: (a) upwind direction; (b) downwind direction; (c) windward direction; (d) leeward direction.  }
	\label{fig:Fig10} 
\end{figure}

It is clear that observing the overlapping parts of different noise spectra in logarithmic coordinates is difficult, and generally it is not necessary to carry out specific analysis of each frequency component. Thus, the 1/3 octave band spectra of sound pressure level were adopted to compare the noise level between the straight-bladed and V-bladed wind turbines. Fig. \ref{fig:Fig10} illustrates the spectra of the sound pressure level of straight blade and V-shaped blades observed at the radial distance of 8$D$ from the center of rotor in four directions at the optimal operating condition $\lambda=2.6$. It can be found that the sound pressure level of all studied wind turbine blades is large in the low-frequency and high-frequency ranges, while the noise level is relatively small in the medium frequency (200 Hz-900 Hz). Besides, it is observed that V-shaped blades generally make less noise in low frequency ($f<100$ Hz). Compared with that, the high-frequency noise of the improved blade is larger than that of baseline model in four directions. On the other hand, the noise emitted from the baseline model is almost the same in different directions (the black solid line in Fig. \ref{fig:Fig10}). On the contrary, the high-frequency noise of V-shaped blades is found to be larger in the windward and leeward directions than that in upwind and downwind directions. This involves the directivity of the VAWT noise propagation, which would be elaborated in the next section.   

\begin{figure}[htbp]
	\centering
	\subfigure[]{
		\begin{minipage}[t]{0.48\linewidth}
			\centering
			\includegraphics[width=1\linewidth]{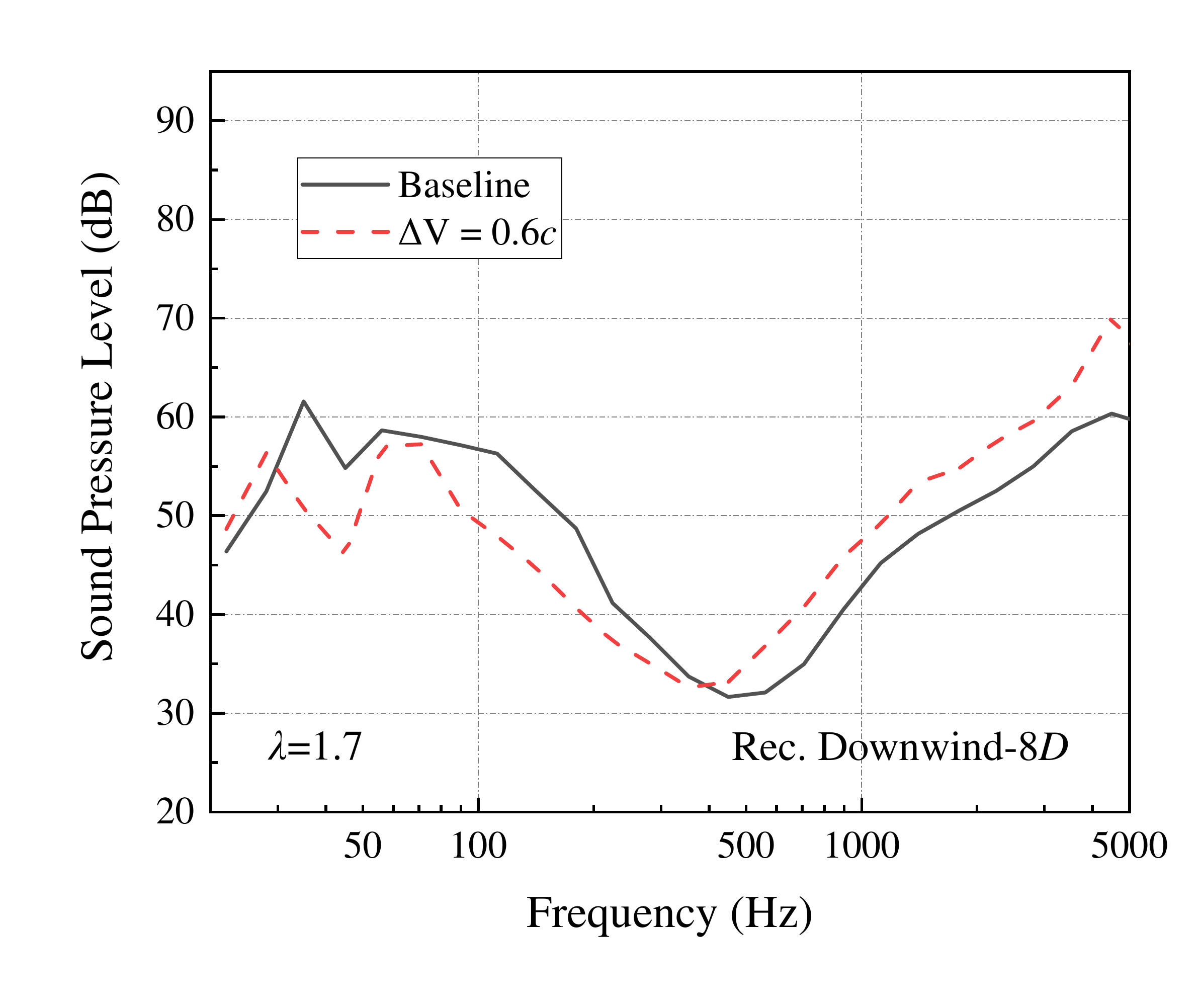}	
	\end{minipage}}	
	\subfigure[]{
		\begin{minipage}[t]{0.48\linewidth}
			\centering
			\includegraphics[width=1\linewidth]{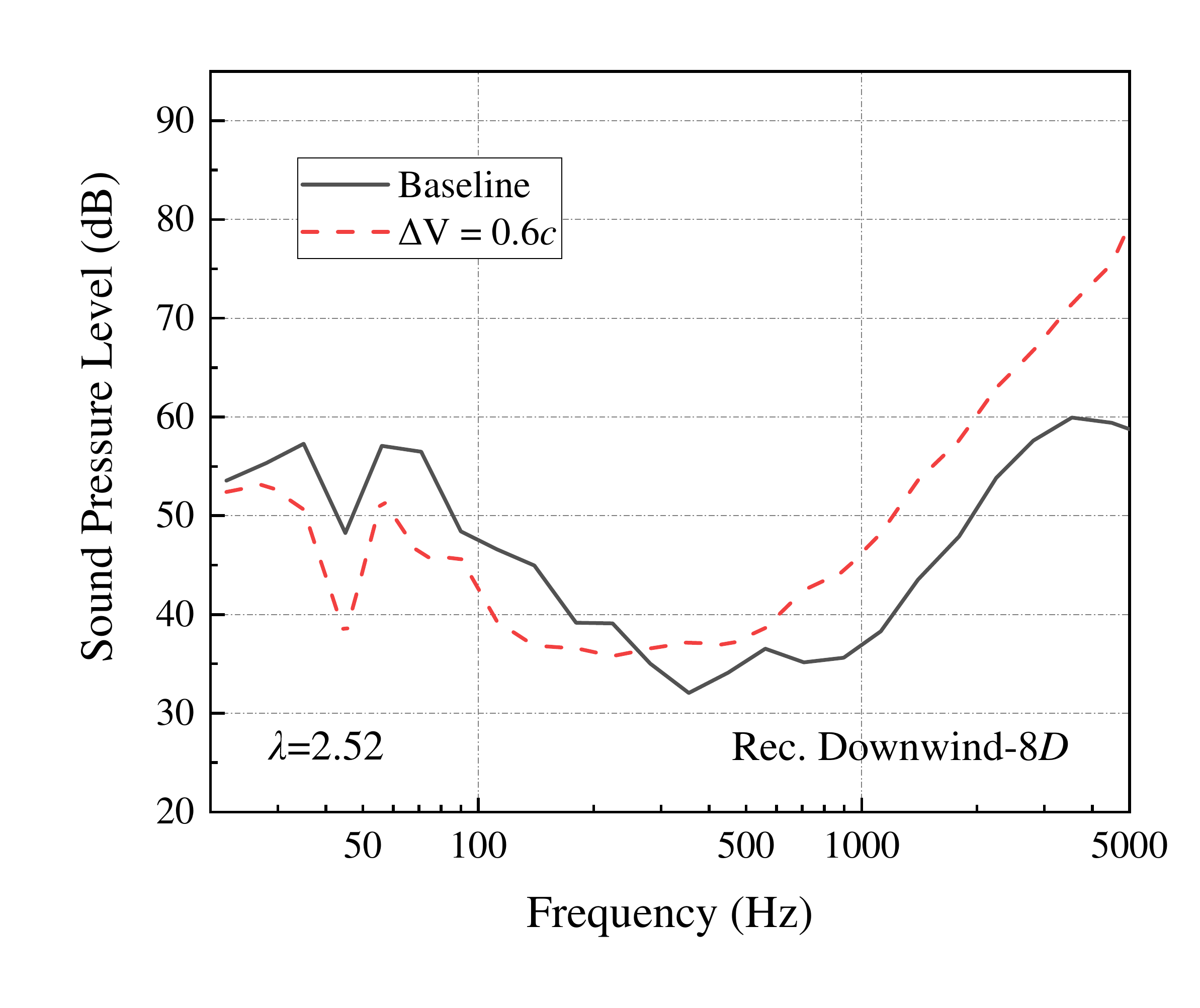}
	\end{minipage}}
	
	\subfigure[]{
		\begin{minipage}[t]{0.48\linewidth}
			\centering
			\includegraphics[width=1\linewidth]{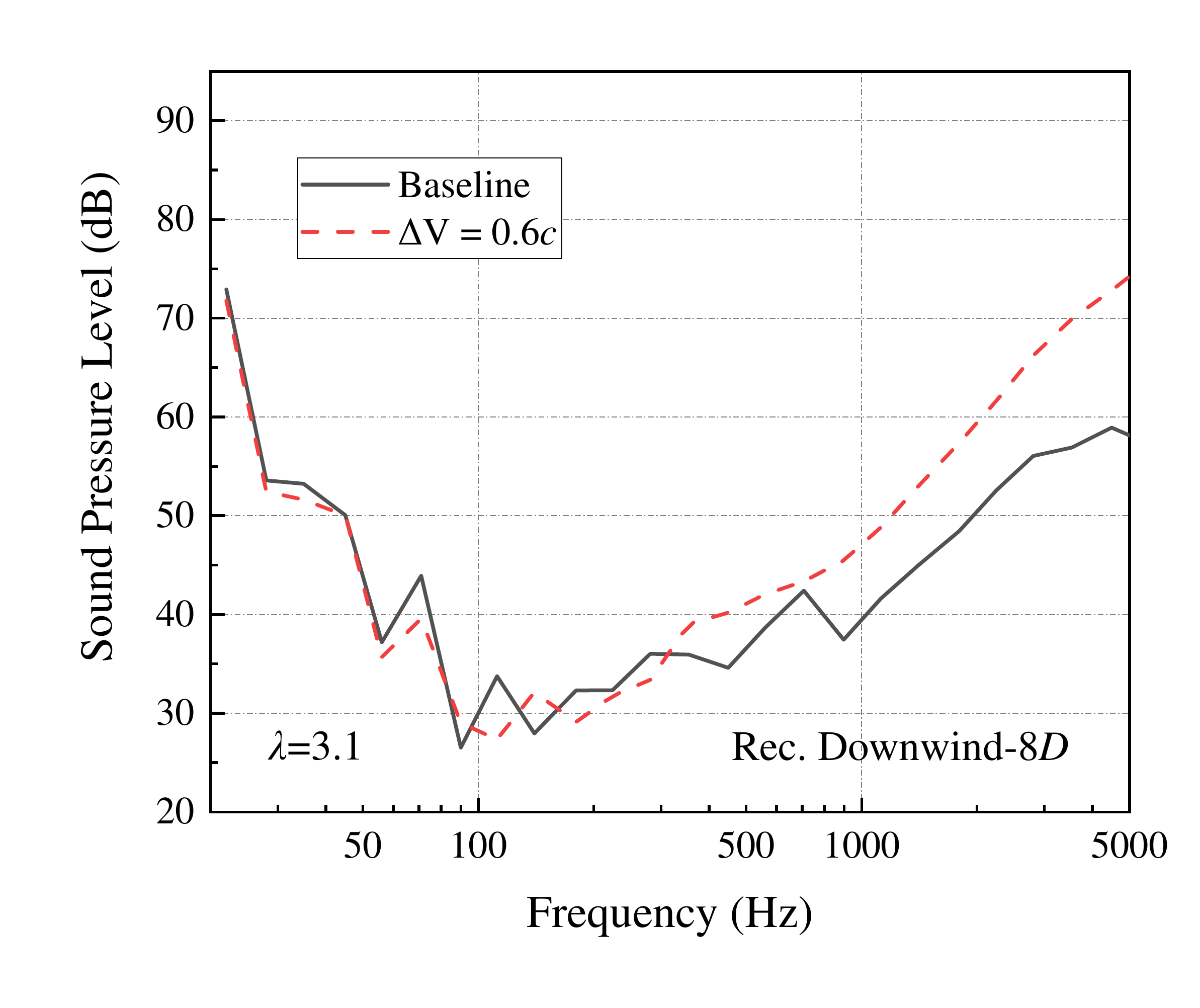}
	\end{minipage}}
	\subfigure[]{
		\begin{minipage}[t]{0.48\linewidth}
			\centering
			\includegraphics[width=1\linewidth]{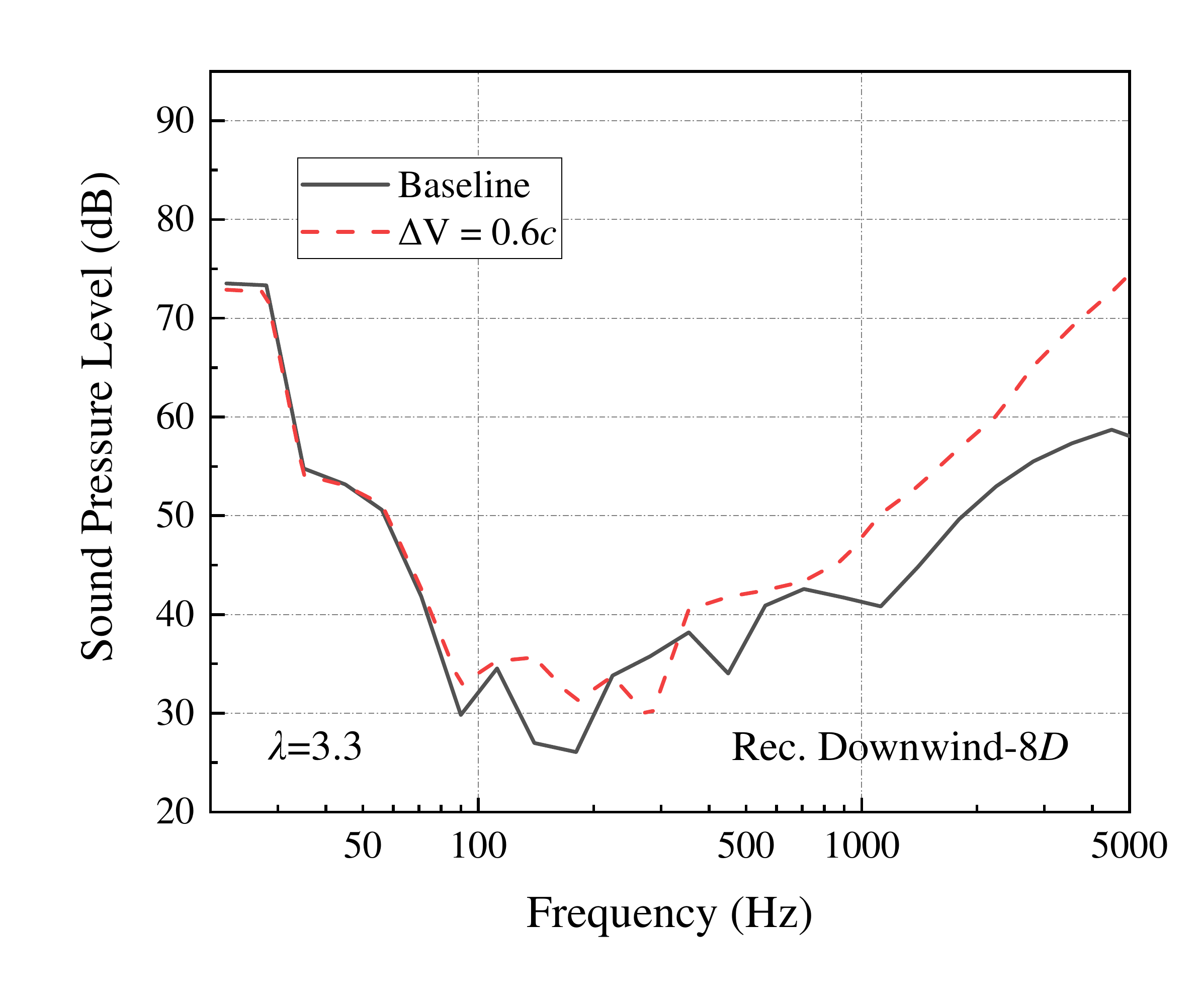}
	\end{minipage}}
	\caption{The 1/3 octave band spectra of the sound pressure level observed at a radial distance of 8$D$ from the center of rotor in the downwind direction at different tip speed ratios: (a) $\lambda=1.7$; (b) $\lambda=2.52$; (c) $\lambda=3.1$; (d) $\lambda=3.3$.  }
	\label{fig:Fig11} 
\end{figure}

For the same reasons as described above, the 1/3 octave band spectra of the V-shaped blade $\Delta V=0.6c$ and the original blade were investigated at other tip speed ratios. The comparison of sound pressure level between straight-bladed wind turbine and V-bladed VAWT at a distance of 8$D$ in the downwind direction is shown in Fig. \ref{fig:Fig11}. The trend that high-frequency noise of V-shaped blade is larger than that of baseline blade also occurs at other four tip speed ratios. In the low frequency range, the noise of V-shaped model is lower. This phenomenon is contrary to the findings on J-shaped blade noise in Ref. \cite{mohamed2019criticism}, where lower noise levels were in high frequencies and high sound levels occurred in low frequencies for modified J-shaped wind turbine. It is known that the low-frequency noise of wind turbines can spread over long distances and easily permeate the building structures, causing high annoyance \cite{liebich2021systematic}. From this perspective, it seems that the V-bladed VAWT that produces lower low-frequency noise performs better for the human body. Besides, it can be found that as the rotational speed increases, the low-frequency range of $SPL_V <SPL_{Base}$ becomes narrow. In general, however, the results suggest that the noise emitted from the V-bladed VAWT is greater. This phenomenon of increasing power while unexpectedly causing increased noise is similar to that of Wind-Lens turbine \cite{hashem2017aero}.

\subsection{Noise directivity}

\begin{figure}[htp]
	\centering
	\subfigure{
		\begin{minipage}[t]{1\linewidth}
			\centering
			\includegraphics[width=1.05\linewidth]{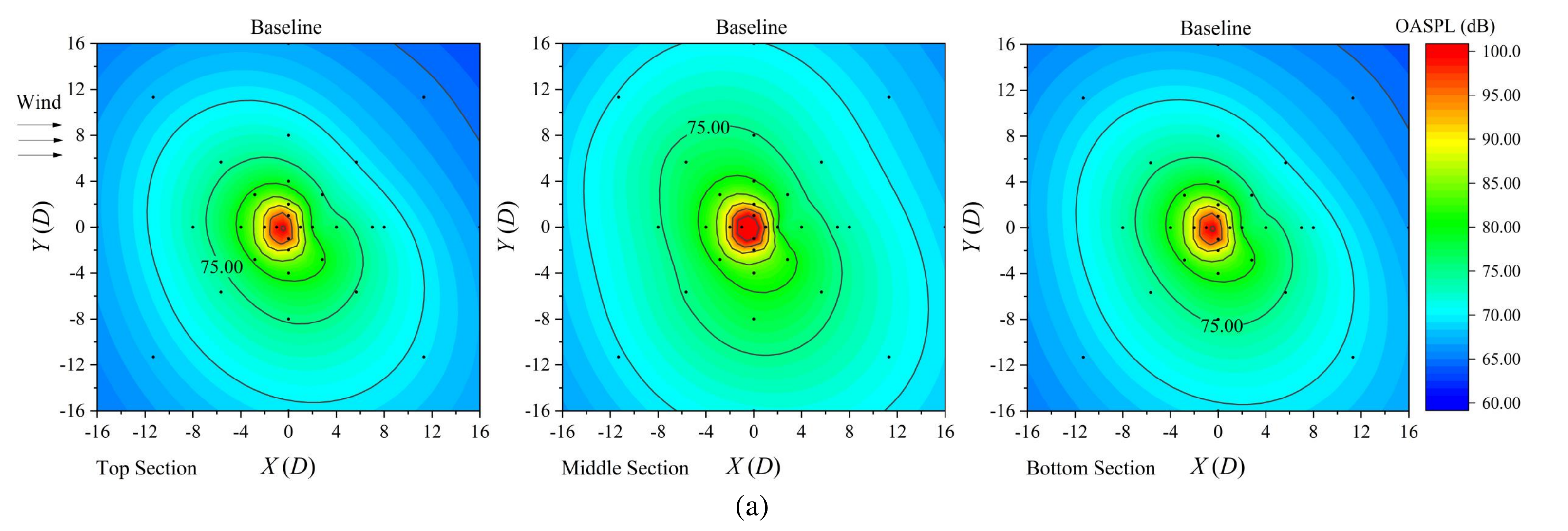}	
	\end{minipage}}
	
	\subfigure{
		\begin{minipage}[t]{1\linewidth}
			\centering
			\includegraphics[width=1.05\linewidth]{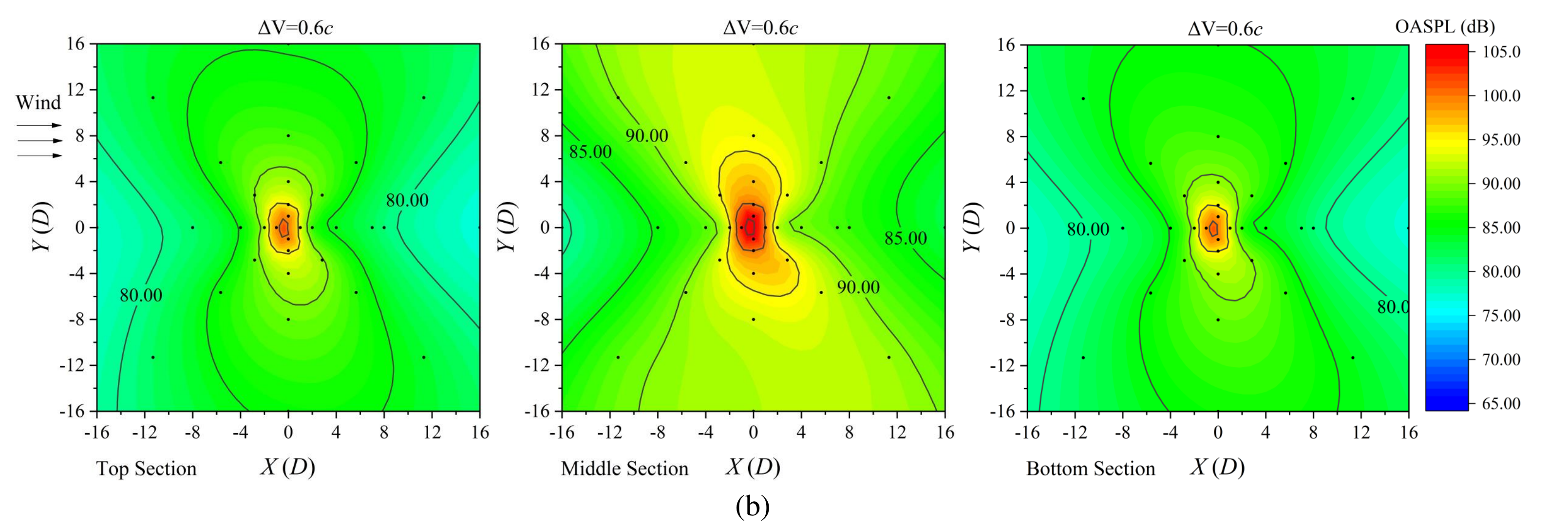}
	\end{minipage}}
	\caption{Directivity and distribution pattern of OASPL of two types of wind turbines at $\lambda=2.6$: (a) straight-bladed VAWT; (b) V-bladed VAWT $\Delta V=0.6c$.  }
	\label{fig:Fig12} 
\end{figure}

The directivity of the predicted overall SPL (OASPL) of two types of wind turbines at three sections with different heights at $\lambda=2.6$ is illustrated in Fig. \ref{fig:Fig12}. For the traditional VAWT, it can be observed that the noise directivity presents a relatively full elliptical distribution characteristic as presented in Fig. \ref{fig:Fig12} (a). The isopleth range of OASPL is slightly smaller in the upwind and downwind directions. This pattern is consistent with the results reported by Aihara et al.\cite{aihara2021aeroacoustic} and Liu et al. \cite{liu2022aerodynamic}, although the data in the former study were only based on the frequency bands from 25 to 3100 Hz. Besides, the noise in the leeward region is found to be larger than that on the windward side, which coincides with the results of Refs. \cite{aihara2021aeroacoustic,liu2022aerodynamic,rasekh2021effect}. On the other hand, an interesting finding in this study is that the V-shaped blades change the distribution pattern of VAWT noise from elliptical to dumbbell-shaped as shown in Fig. \ref{fig:Fig12} (b). The noise level in the windward and leeward regions is obviously larger than that in the upwind and downwind directions. Besides, it can be observed that the noise level at the middle horizontal section is higher and the radiation range is wider. It is found that at the same distance from the center of wind turbine, the noise level at the middle section is about 5 dB higher than that at the top or bottom section. It suggests that strategically altering the height position of small VAWTs may cause less noise annoyance for potential receivers in urban areas. Based on the previous study, the enhancement in power coefficient was mainly attributed to the middle part of the V-shaped blade, while more energy was lost in the blade tip region \cite{su2020investigation}. Besides, a pair of symmetric vertical structures formed near the blade midspan due to the induced downward velocity \cite{zhang2020laminar}. This uneven distribution of blade surface loads as well as the pressure fluctuations may partly explain the larger noise emitted from V-bladed VAWT.

\begin{figure}[htp]
	\centering
	\includegraphics[width=0.4\linewidth]{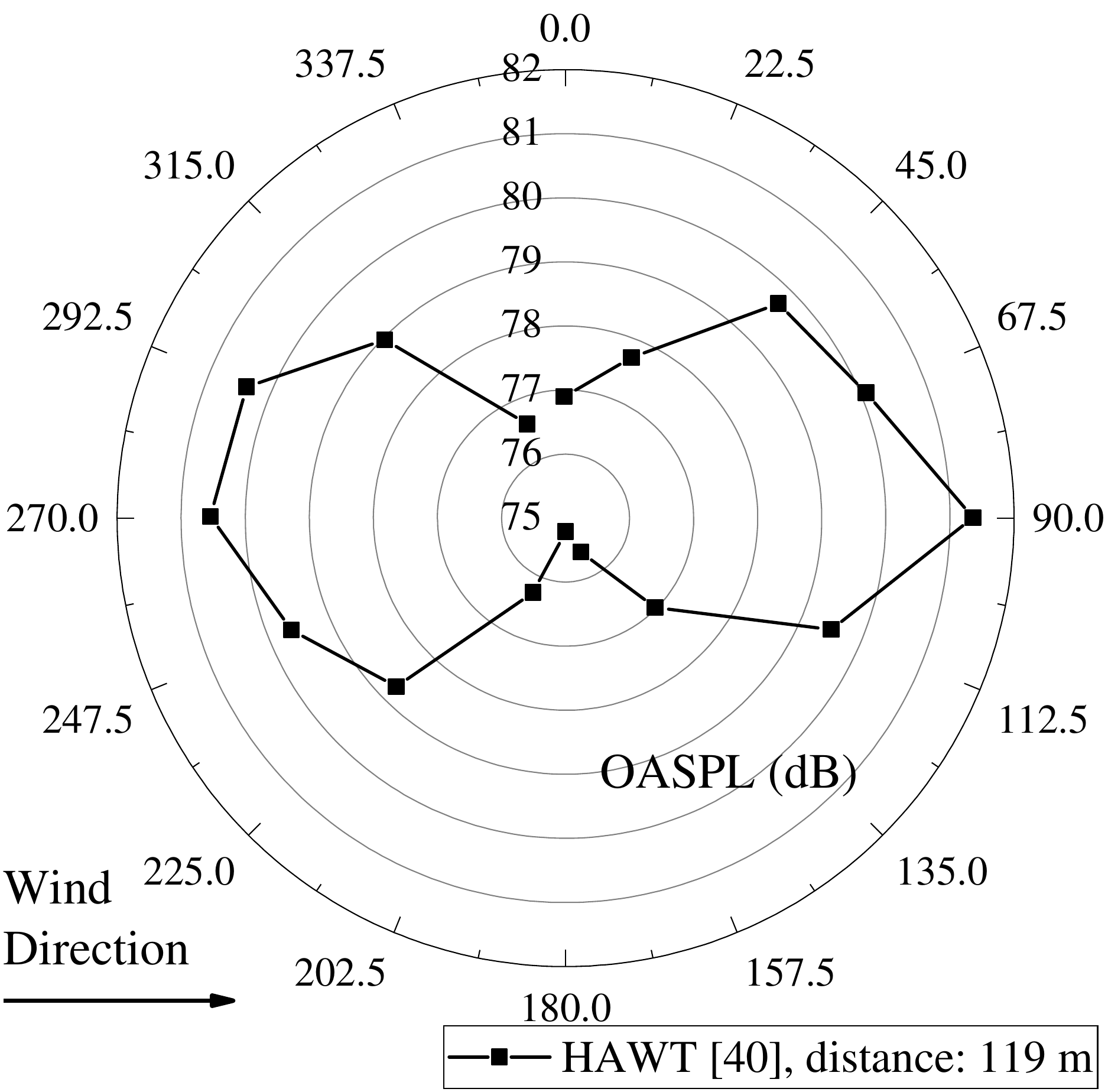}
	\caption{ The noise directivity of the HAWT \cite{kaviani2017aerodynamic}.  }
	\label{fig:Fig13} 
\end{figure}

It is well known that the rotor of the HAWT is a typical dipole sound source. Thus, the noise directivity of HAWTs usually presents a dumbbell-shaped distribution as shown in Fig. \ref{fig:Fig13} \cite{kaviani2017aerodynamic}, or has a symmetrical cardioid shape as revealed the BPM method in Refs. \cite{zhu2005modeling,zhu2018wind,bresciani2022influence}. Based on the similarity of distribution patterns between the V-bladed VAWT and conventional HAWTs, it can be inferred that the effect of dipole noise, namely the loading noise, becomes larger for the V-bladed VAWT. Besides, it should be noted that the orientation of dumbbell-shaped distribution pattern of the V-bladed VAWT is different from that of the HAWT. With the wind blowing from left to right, as shown in Fig. \ref{fig:Fig12} (b) and Fig. \ref{fig:Fig13}, the two ends of the dumbbell pattern of the V-bladed VAWT are in the windward and leeward regions, while the two ends of the dumbbell pattern of the HAWT are in the upwind and downwind positions, respectively. This dumbbell-shaped distribution could be explored for optimizing wind farms to minimize noise disturbance to potential residents.

\begin{figure}[htp]
	\centering
	\subfigure{
		\begin{minipage}[t]{1\linewidth}
			\centering
			\includegraphics[width=0.9\linewidth]{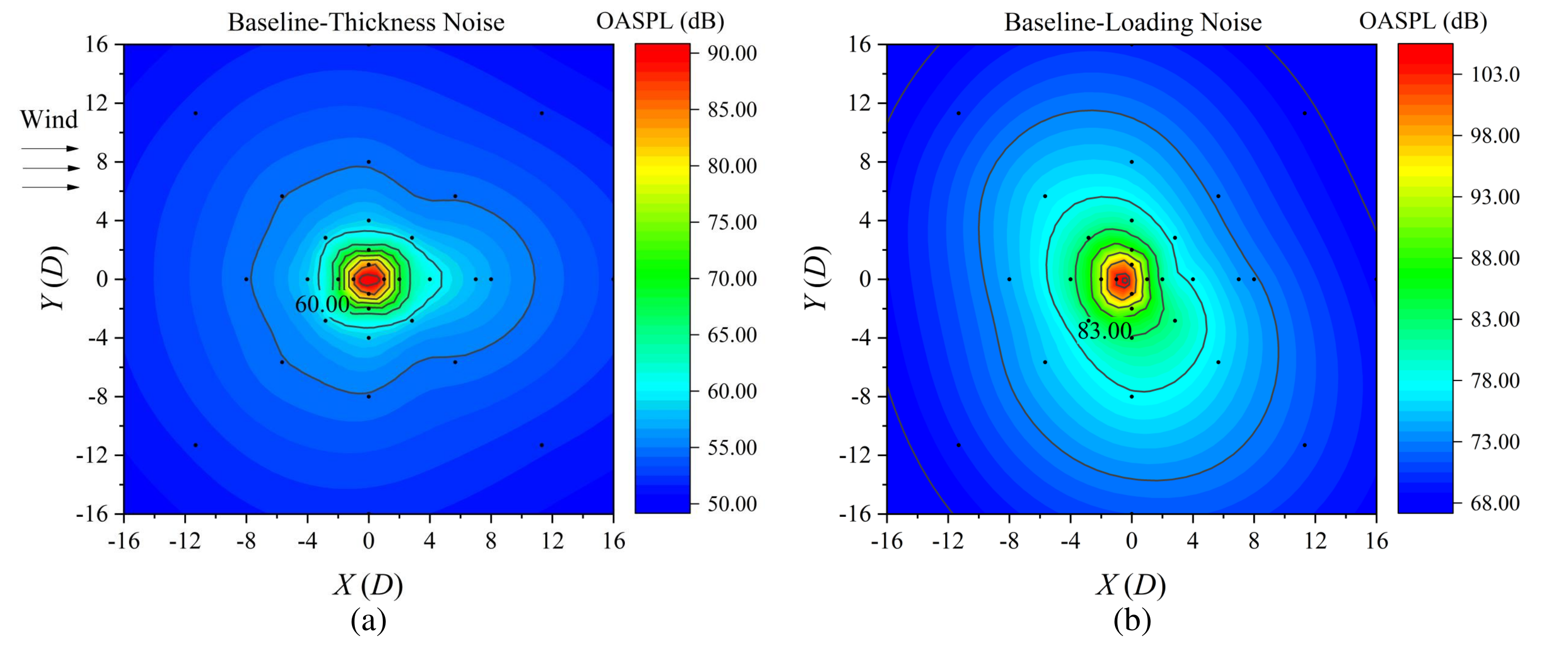}	
	\end{minipage}}
	\subfigure{
		\begin{minipage}[t]{1\linewidth}
			\centering
			\includegraphics[width=0.9\linewidth]{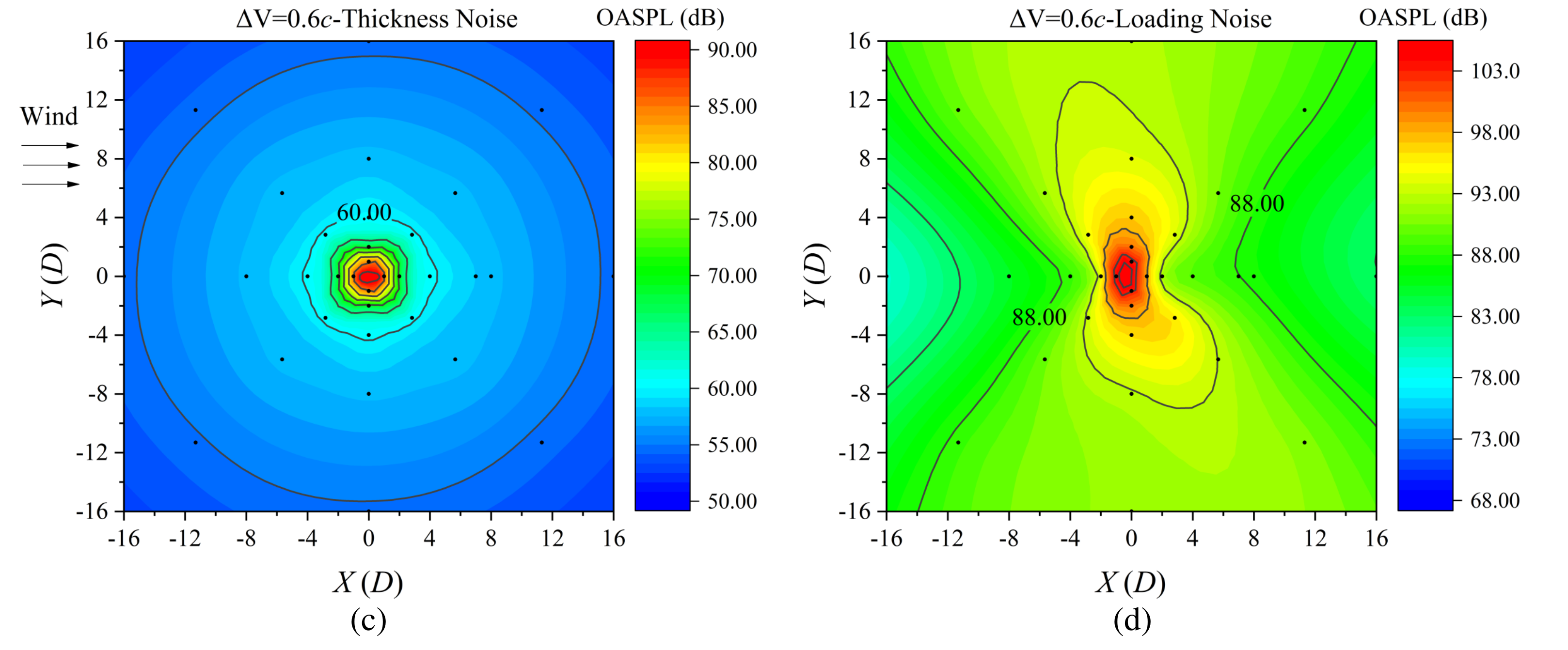}
	\end{minipage}}
	\caption{Comparison of the components of total noise:  (a) the thickness noise of baseline model; (b) the loading noise of baseline model; (c) the thickness noise of V-bladed VAWT $\Delta V=0.6c$; (d) the loading noise of V-bladed VAWT $\Delta V=0.6c$.  }
	\label{fig:Fig14} 
\end{figure}

To make further investigation on the physics behind the shift in the noise directivity patterns, the components of total noise are studied. As is well known, the tonal noise emitted from VAWTs can be divided into the thickness, loading, and quadrupole noise, among which the thickness and loading noise are the dominant noise sources while the quadrupole noise usually has negligible influence. Therefore, the distribution patterns of thickness and loading noise of the baseline model and V-bladed turbine were investigated as presented in Fig. \ref{fig:Fig14}. It is clear that the thickness noise, as also called the monopole noise, is nearly uniform in directivity in all directions. The V-shaped blade increases the noise level of thickness noise, but the distribution pattern does not change. On the other hand, in the case of applying V-shaped blades, the characteristics of the dipole sound source are more pronounced. As shown in Fig. \ref{fig:Fig14} (d), the dumbbell-shaped distribution is clearer, and the propagation range of the loading noise is larger.

Moreover, the overall noise level of the studied VAWTs was compared with that of the HAWTs with similar size as shown in Table \ref{table 4}. Here, the results of the noise level of HAWTs were published by Ocker and Blumendeller et al. \cite{ocker2022localization}. The reason for choosing this set of noise data for HAWTs is that the sizes of the tested HAWTs are similar to that of the present VAWT, and the operating conditions are basically the same. Besides, the Strouhal number and Mach number of the scaled HAWTs in the experiments are comparable to those of full-size wind turbines, which means that to some extent it can reflect the acoustic properties of large HAWTs. The comparison of the noise levels between the studied VAWTs and the HAWTs with similar size is presented in Table \ref{table 4}. It can be observed that the noise levels of VAWTs are generally lower than those of HAWTs. Even if the V-shaped structure increase the noise emitted from the VAWT, its OASPL is still lower than that of HAWT under similar operating conditions.

\begin{table}[h]
	\centering
	\caption{Comparison of the overall SPL between the studied VAWTs and the HAWTs with similar size.}
	\begin{tabular}{c | c c c | c c c }
		\toprule
		\multirow{2}{*}{WT}&Receiver &\multirow{2}{*}{Diameter} &\multirow{2}{*}{Blade}&\multicolumn{3}{c}{OASPL (dB)}\\	
		\cline{5-7}
		&distance&&&$U_{\infty}$=5 m/s&7 m/s&9 m/s	\\
		\hline
		\multirow{3}{*}{HAWT\cite{ocker2022localization}}&\multirow{3}{*}{1.96 m}&1.5 m&NACA 4412 shape&101.2&107.7&111.9\\
		&&1.29 m&Clark-Y shape&102.4&104.4&114.4\\
		&&1.16 m&Sickle shape&101.0&106.8&111.5\\
		
		\midrule[1.pt]
		
		\multirow{4}{*}{VAWT}&Receiver &Diameter &\multirow{2}{*}{Blade}&\multicolumn{3}{c}{OASPL (dB)}\\	
		\cline{5-7}
		&distance&and height&&\multicolumn{3}{c}{$U_{\infty}$=9 m/s}	\\
		\cline{2-7}
		&\multirow{2}{*}{2.1 m}&1.05m$\times$&NACA0021
		Straight
		&\multicolumn{3}{c}{79.6}\\
		&&1.465m&NACA0021 V-shaped
		&\multicolumn{3}{c}{91.8}\\
		
		\bottomrule
	\end{tabular}	
	
	\label{table 4}
\end{table}

\subsection{Effect of trailing-edge serrations}

The previous results indicate that the V-shaped blade could effectively increase the power output of a VAWT, although its OASPL is higher than that of the conventional straight-bladed VAWT. It suggests that there is value in reducing noise of the V-bladed VAWT while increasing power performance as much as possible. Therefore, this subsection attempts to further explore the noise reduction effect of trailing-edge serrations on the V-bladed wind turbine.

\subsubsection{Aerodyamics}

\begin{figure}[htbp]
	\centering
	\subfigure[]{
		\begin{minipage}[t]{0.485\linewidth}
			\centering
			\includegraphics[width=1\linewidth]{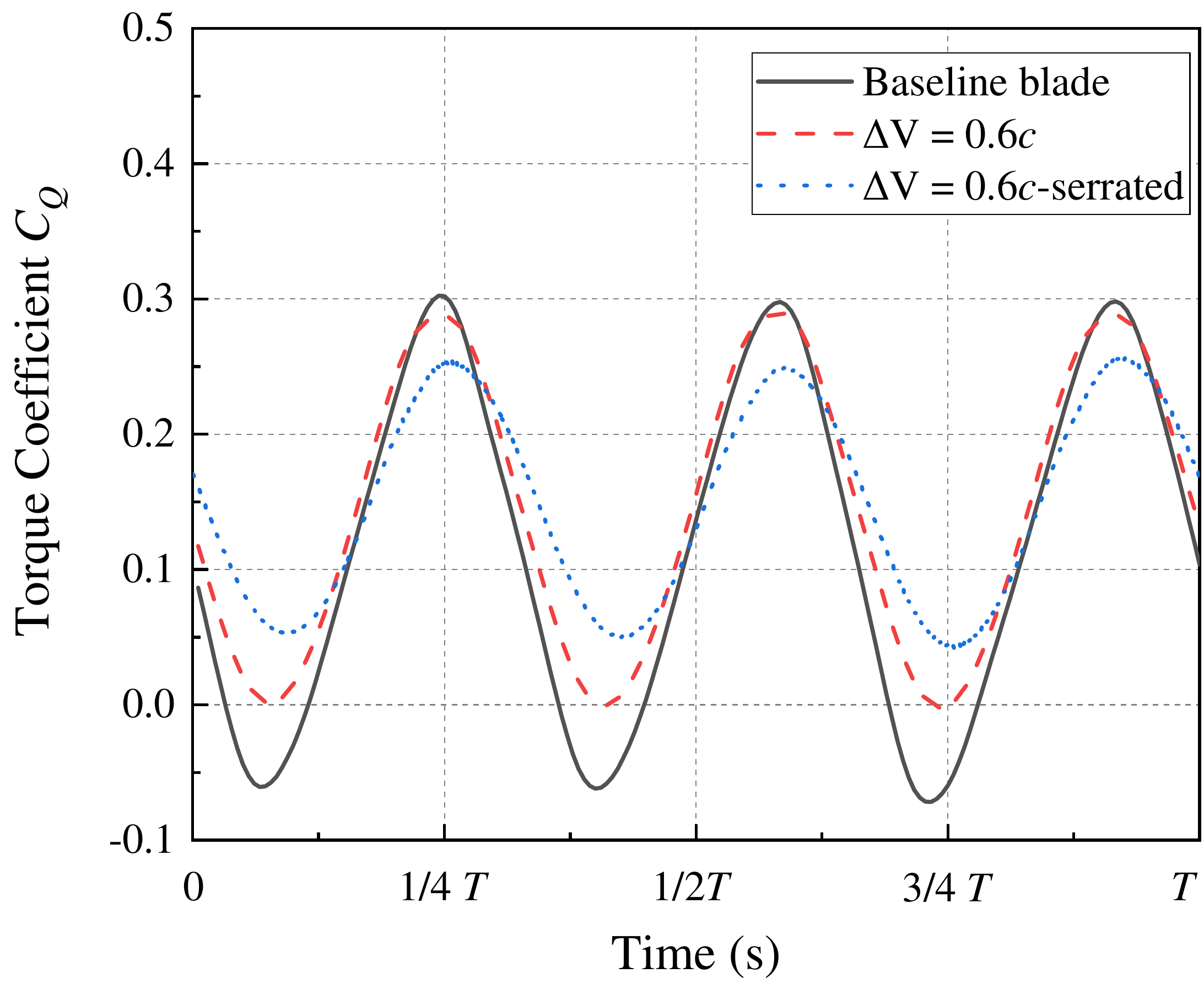}	
	\end{minipage}}	
	\subfigure[]{
		\begin{minipage}[t]{0.485\linewidth}
			\centering
			\includegraphics[width=1\linewidth]{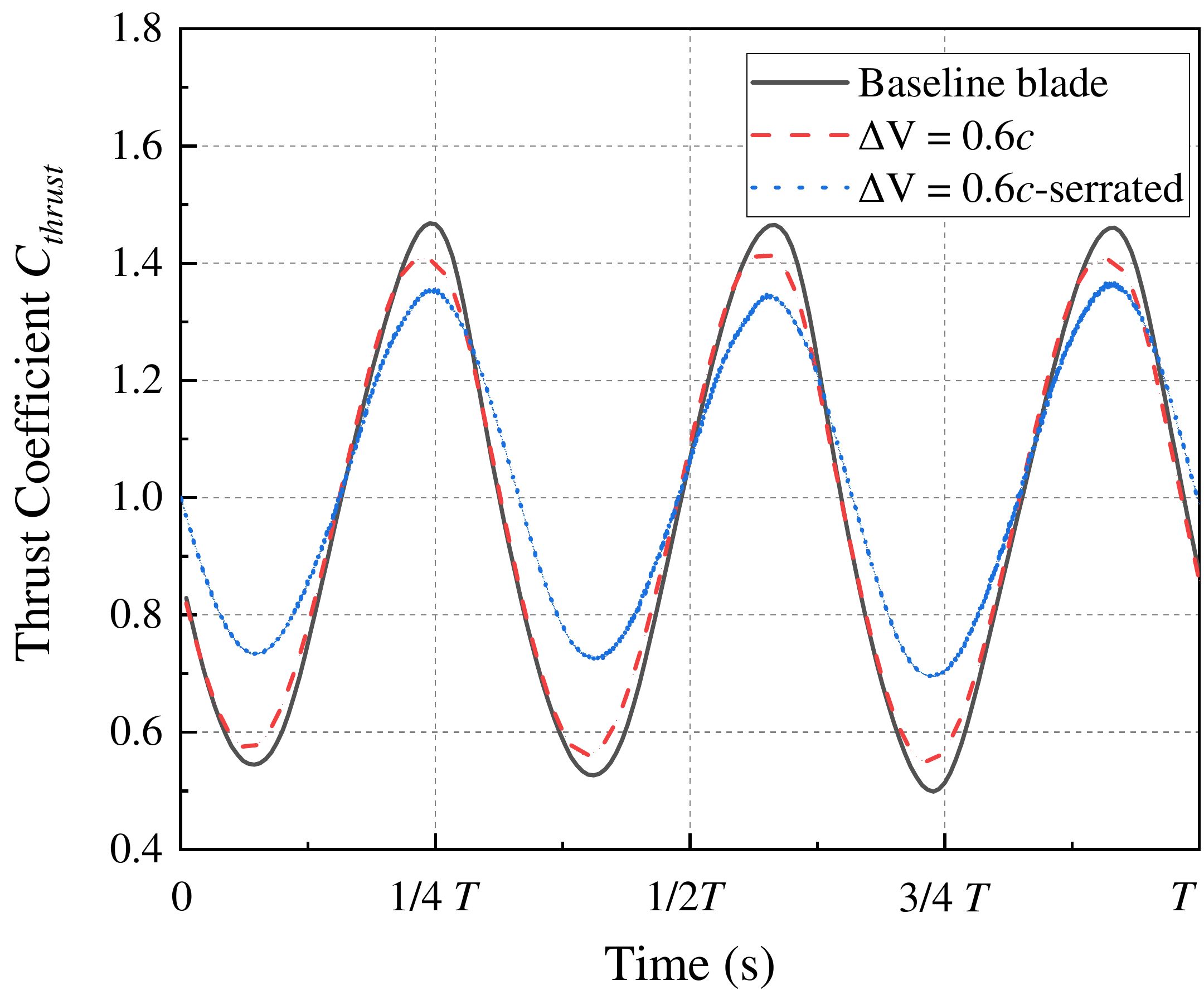}
	\end{minipage}}
	
	\caption{Comparison of the fluctuations of the torque coefficient and thrust coefficient of the whole rotor with different blades: (a) torque coefficient; (b) thrust coefficient.
	}
	\label{fig:Fig8} 
\end{figure}

The variations of torque coefficient and thrust coefficient for the VAWT with different modified blades at the optimal tip speed ratio $\lambda=2.6$ are shown in Fig. \ref{fig:Fig8}. Besides, the averaged values of torque coefficient $C_{Q}$, power coefficient $C_{P}$ and thrust coefficient $C_{thrust}$ are calculated as listed in Table \ref{table 3}. It can be observed that the output torque coefficient $C_{Q}$ for the baseline blade fluctuates most violently, where the peak value of $C_{Q}$ exceeds 0.3 and the lowest value is less than 0. Compared with that, the peak value of $C_{Q}$ of the V-bladed wind turbine is slightly less than that of the straight-bladed model, while the lowest $C_{Q}$ is greatly improved by the V-shaped blades due to its superiority in reducing drag coefficient. In this way, the V-bladed VAWT can achieve higher power output with up to 24.11$\%$ increase in $C_{Q}$ and $C_{P}$. After applying the trailing-edge serrations, the peak value of $C_{Q}$ of the wind turbine rotor decreases, and the fluctuations are further reduced. The results show that the averaged torque output of the modified wind turbine blade with trailing-edge serrations is slightly higher ($3.35\%$) than that of the blade without this structure. For the thrust coefficient, similarly, the fluctuation of $C_{thrust}$ for the V-shaped blade with trailing-edge serrations is smallest, and the averaged $C_{thrust}$ is also higher than that without trailing-edge serrations. 

\begin{table}[h]
	\caption{The averaged torque coefficient $C_{Q}$, power coefficient $C_{P}$ and thrust coefficient $C_{thrust}$ for the VAWT with different modified blades at $\lambda=2.6$.}
	\centering
	\begin{tabular}{l l l c l c}
		\toprule
		Blade& $C_{Q}$&$C_{P}$& Increment ($\%$)& $C_{thrust}$&Increment ($\%$) \\
		\hline
		Baseline&0.1162 & 0.3023 & -&0.9766&-\\
		$\Delta V=0.6c$&0.1443 & 0.3751 & 24.11&0.9809&0.44\\
		$\Delta V=0.6c$ -serrated &0.1491 & 0.3877 & 28.27&1.0297&5.43\\
		\bottomrule
	\end{tabular}
	\label{table 3}
\end{table}

In the previous studies, it was generally considered that utilizing trailing-edge serrations or sawtooth structures would deteriorate the aerodynamic performance of the single blade \cite{liu2017aerodynamic,llorente2019trailing}. However, it was also reported that this kind of trailing-edge modification would slightly increase the power performance of wind turbines under certain conditions \cite{llorente2020trailing,zhou2022performance}. The present results confirm this phenomenon on the VAWT, although it has only been found on HAWTs before \cite{llorente2020trailing}. To explain this difference in the effect of trailing edge serration structure on blade performance under different conditions, more research is needed in the future.

\subsubsection{Aeroacoustics}

\begin{figure}[htp]
	\centering
	\subfigure{
		\begin{minipage}[t]{1\linewidth}
			\centering
			\includegraphics[width=1.05\linewidth]{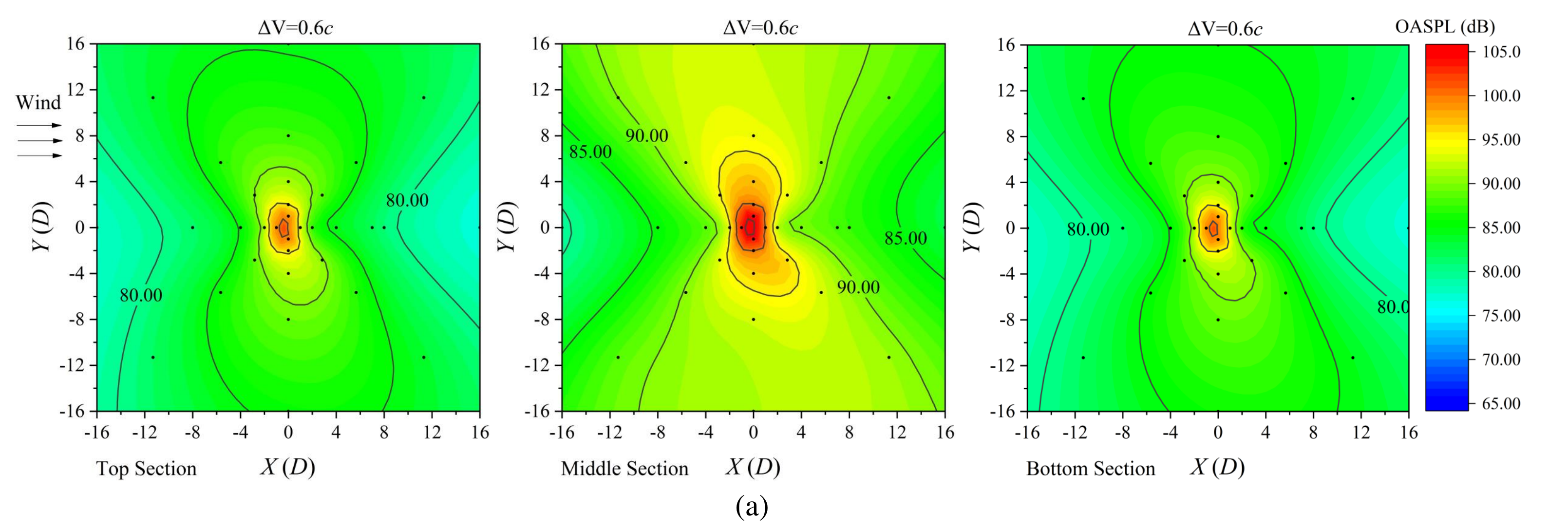}	
	\end{minipage}}
	
	\subfigure{
		\begin{minipage}[t]{1\linewidth}
			\centering
			\includegraphics[width=1.05\linewidth]{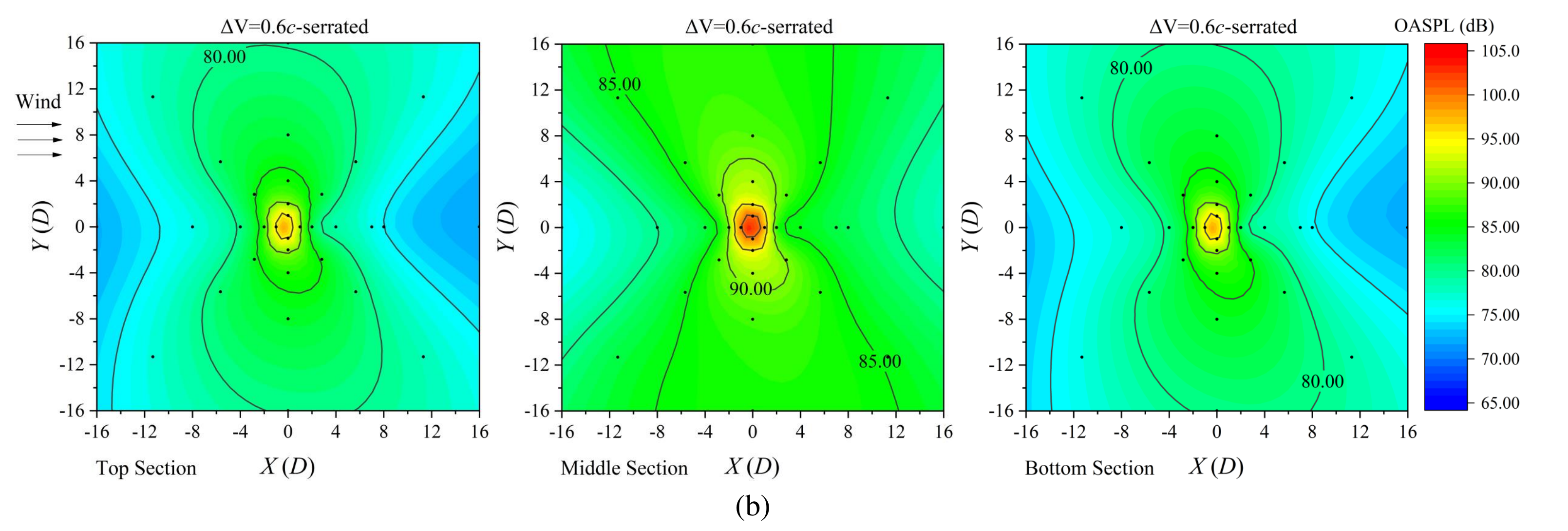}
	\end{minipage}}
	\caption{Directivity and distribution pattern of OASPL of two types of wind turbines at $\lambda=2.6$: (a) V-bladed VAWT $\Delta V=0.6c$; (b) V-bladed VAWT $\Delta V=0.6c$ with trailing-edge serrations.  }
	\label{fig:Fig15} 
\end{figure}

Fig. \ref{fig:Fig15} presents the comparison of the tonal noise between V-shaped blades and V-shaped blades with trailing-edge serrations at different heights. The overall SPL over the entire frequency range was used to depict the distribution pattern of wind turbine noise. It can be observed that applying the serrations to the blade trailing edge would reduce the noise emitted from the wind turbine. The noise reduction mechanism of trailing-edge serration structure has been extensively studied \cite{azarpeyvand2013analytical,gruber2012airfoil}. The previous results suggested that the noise reduction due to serration structure originated primarily from the interference effects near the blade trailing edge \cite{gruber2012airfoil}. Besides, the use of serrations can lead to significant reduction of turbulent kinetic energy and Reynolds shear stress in the near wake region, thereby reducing the turbulence interaction noise \cite{liu2016wake}. According to the values marked on the isopleth in Fig. \ref{fig:Fig15}, the OASPL of serrated V-bladed wind turbine is about 5 dB lower than that of the unmodified V-bladed VAWT at the same distance from the center of wind turbine. But it is clear that the dumbbell-shaped noise distribution pattern has not changed. The results indicate that the trailing-edge serrations could reduce the noise of VAWTs.

\begin{figure}[htp]
	\centering
	\includegraphics[width=0.5\linewidth]{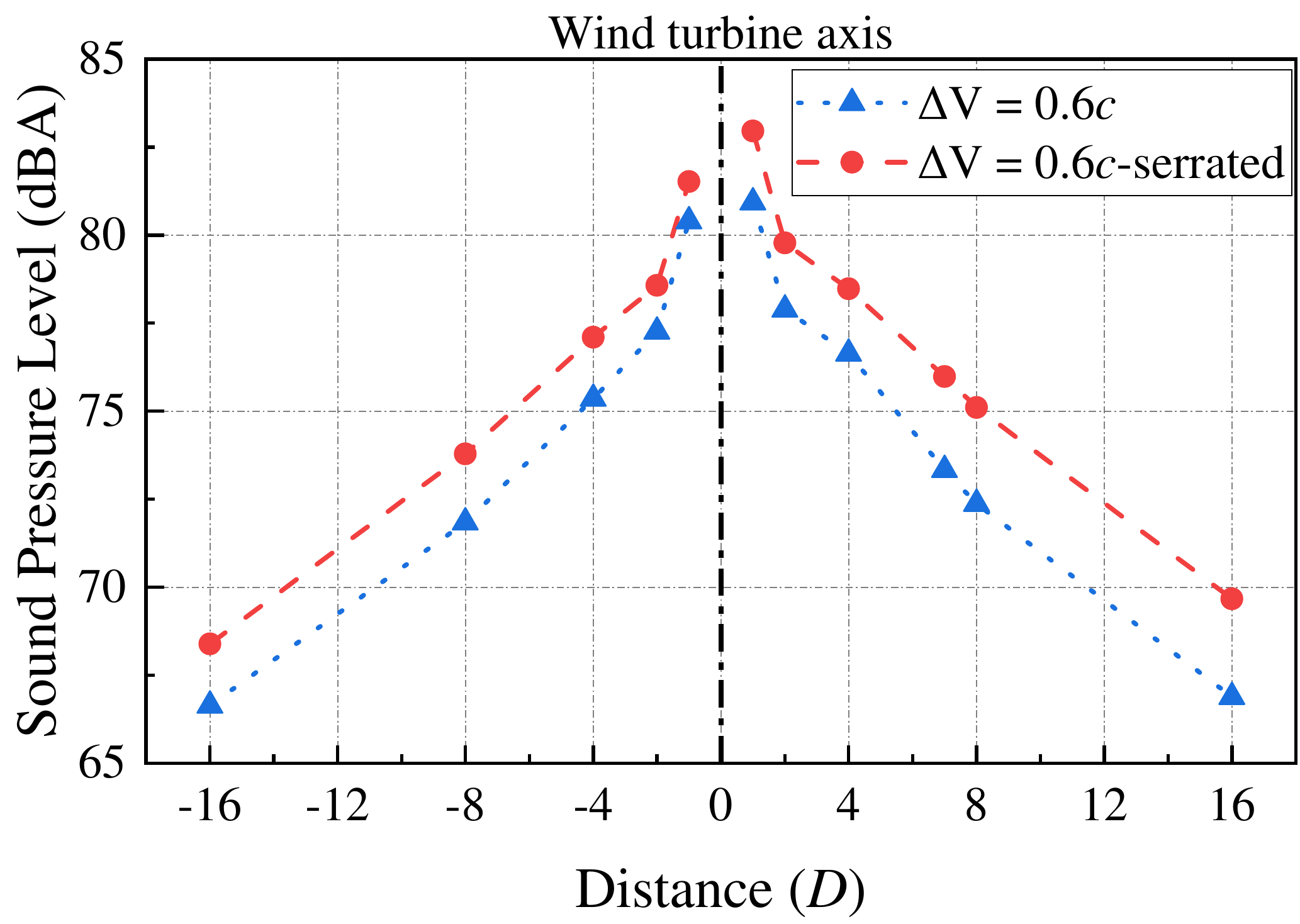}
	\caption{The recorded A-weighted OASPL at different distances for the studied wind turbine with different blades.}
	\label{fig:Fig16}
\end{figure}

To better assess the noise level to which human hearing is sensitive, as well as to quantify the noise reduction effect of wind turbines, the A-weighted overall SPL based on the frequency bands from 25 to 5000 Hz was calculated. Fig. \ref{fig:Fig16} shows the A-weighted OASPL of the studied wind turbine with different blades. It can be found that the noise level decreases with the distance from the wind turbine center. The A-weighted noise level of V-bladed wind turbine is larger than that of the straight-bladed turbine. After applying the trailing-edge serrations, the noise level is reduced by up to 2.8 dBA, and the average noise reduction is about 2.3 dBA. This effect is similar to that found in Ref. \cite{oerlemans2009reduction}. Because this section is a preliminary application study of trailing-edge serrations on the V-bladed VAWT, only one recommended serration structure was investigated. The optimal trailing-edge serrations for the VAWT with V-shaped blades will be studied in the future work. In summary, the power and thrust coefficients of the studied V-bladed VAWT with trailing-edge serrations are about 28.3$\%$ and 5.4$\%$ higher than those of the baseline turbine. Although the A-weighted noise level of the former is higher than that of the original wind turbine, its noise level is still much lower than that of the HAWT. 

\section{Conclusion and outlook}

The present study investigated the aerodynamic performance and aeroacoustic characteristics of a small VAWT with V-shaped blades under different rotational speeds. The CFD model combined with FW-H method was utilized to evaluate the effects of different V-shaped blades on the aerodynamic forces and the wind turbine noise level. In addition, as one of the noise reduction techniques, the trailing-edge serration structure was applied to the wind turbine blades to assess the noise reduction effect. The main conclusions are summarized as follows:

1. At the optimal tip speed ratio $\lambda=2.6$ of the VAWT, the V-shaped blade $\Delta V=0.6c$ achieved the highest enhancement in power output compared with the other blade models. This type of V-shaped blade could effectively improve the power performance of VAWT over a wide range of tip speed ratios. The maximum increase in power coefficient can reach about 24.1$\%$, while the average power increment for the improved blade is about 14.6$\%$ from moderate to high tip speed ratios.

2. Under all operating conditions, the V-shaped blades only slightly change the thrust coefficient of the VAWT. Specifically, the V-bladed wind turbine generates lower thrust coefficient at low rotational speeds, and produces larger value of $C_{thrust}$ at high tip speed ratios. The increase in $C_{thrust}$ on the structure is less than 4$\%$.

3. The trailing-edge serrations slightly increase the rotor torque output and thrust of V-bladed VAWT. The power and thrust coefficients of the V-bladed wind turbine with trailing-edge serrations are about 28.3$\%$ and 5.4$\%$ higher than those of the baseline turbine at the optimal tip speed ratio.

4. In the acoustic spectrum analysis, the V-bladed VAWT generates less low-frequency noise compared with that of the original model in the four noise propagation directions under different tip speed ratios, which is better for the human body. As the rotational speed increases, the low-frequency range of $SPL_V <SPL_{Base}$ became narrow.

5. The noise directivity of the straight-bladed VAWT presented a relatively full elliptical distribution characteristic, while a dumbbell-shaped noise directivity distribution was discovered for the studied V-bladed wind turbine, which can be utilized to reduce noise disturbance in wind farm layout.

6. The trailing-edge serrations realized the expected noise reduction effect, and the average noise reduction was about 2.3 dBA. Although the A-weighted noise level of the former is higher than that of the original wind turbine, its noise level is still much lower than that of the HAWT. 

In the practical engineering applications, the cost and difficulty of making V-shaped blades could be reduced by utilizing the existing manufacturing process of straight blades. The trailing-edge serrations can be made separately and installed on the blade as an optional part to meet the requirements of different work scenarios. 

In summary, the results of this work add new knowledge to the design of VAWTs with high power output and low noise annoyance. However, there are still some limitations in this study. In the future, research should focus on the structural response of V-bladed wind turbine, the parameter optimization of trailing edge serrations and their distribution along the blade span. Besides, the mechanism that the trailing edge serrations can improve the power output under certain conditions is also worth exploring

\section*{Acknowledgments}
The authors gratefully acknowledge the financial support by the Innovation Program of Shanghai Municipal Education Commission (No. 2019-01-07-00-02-E00066), the National Natural Science Foundation of China (Nos. 42076210, 52122110, 52088102, 12202271) , and Oceanic Interdisciplinary Program of Shanghai Jiao Tong University (Nos. SL2020PT201, SL2021PT302).

\bibliographystyle{unsrtnat}
\bibliography{references}  






\end{document}